\documentclass[onecolumn,aps,english,a4paper,floatfix]{revtex4}

\usepackage[T1]{fontenc}
\usepackage{tikz}
\usepackage{color}
\usepackage{amsmath}
\usepackage{amssymb}
\usepackage[latin1]{inputenc}
\usepackage{graphicx}
\usepackage{bm}
\usepackage{epsfig}
\usepackage{xcolor}
\definecolor{mynicegreen}{RGB}{102,182,102}

\begin{document}


\title{Exotic liquid crystalline phases in monolayers of vertically vibrated granular particles}

\author{
	Y. Mart\'{\i}nez-Rat\'on\textsuperscript{a} and E. Velasco\textsuperscript{b}}
\affiliation{\textsuperscript{a}
Grupo Interdisciplinar de Sistemas Complejos (GISC), Departamento
de Matem\'aticas, Escuela Polit\'ecnica Superior, Universidad Carlos III de Madrid,
Avenida de la Universidad 30, E-28911, Legan\'es, Madrid, Spain\\
\textsuperscript{b}
Departamento de F\'{\i}sica Te\'orica de la Materia Condensada,
Instituto de F\'{\i}sica de la Materia Condensada (IFIMAC) and Instituto de Ciencia de
Materiales Nicol\'as Cabrera,
Universidad Aut\'onoma de Madrid,
E-28049, Madrid, Spain
}


\begin{abstract}
Vibrated monolayers of granular particles confined into horizontal cavities form a variety of fluid patterns with orientational order that resemble equilibrium liquid-crystal phases. In some cases one can identify nematic and smectic patters that can be understood in terms of classical statistical mechanics of hard bodies. Low aspect ratio cylinders project as rectangles and form uniaxial, or 2-atic, and tetratic, or 4-atic, nematic phases. Other polygonal particles may exhibit different liquid-crystal phases, in general $p$-atic phases, of higher symmetries. We give a brief summary of theoretical work on rectangles and triangles, and provide some experimental results on vibrated monolayers. In the case of equilateral triangles, the theory predicts an exotic triatic phase, or 6-atic phase, with six-fold symmetry and three equivalent directors. Right-angled triangles exhibit a 4-atic phase with strong octatic (8-atic) correlations. Experiments on cylinders show 4-atic textures and, even more remarkable, geometric frustration caused by confinement excites topological defects, which seem to follow the same topological rules as standard liquid crystals. Some of our findings can be understood with the help of simulations of hard particles subject to thermal equilibrium, although standard Density-Functional Theories fail to account for the correct equilibrium phases in some cases. 
\end{abstract}


\maketitle

\section{Introduction}

Hard particles continue to generate a lot of interest in condensed-matter physics, not only because of 
the potential technological applications based on their self-assembling properties, but also because 
the prominent role that entropy plays in their behaviour. 
Recent advances in the synthesis of particles have
provided the possibility to produce particles with a fine control of their size, shape and interactions
(see e.g. \cite{Zhao0,Zhao1,Zhao2}), opening an enormous field of experimental research concerning phase
transitions involving positional and orientational degrees of freedom.
By suitable treatment these interactions can be chosen to be
very close to hard-core interactions. In solution these particles can be considered as
quasi-hard particles subject to Brownian motion and theoretical analyses based on the 
traditional density-functional theories (DFT) for hard bodies are appropriate.
The variety of particles shapes and the subtle entropic forces that operate in these systems
have been shown to produce new possibilities for the design of self-assembling materials 
\cite{reviewGlotzer,Glotzer0,Dijkstra0} and therefore theoretical studies exploring the fundamental ordering
mechanisms have been revitalised.

Recently two-dimensional (2D) hard particles have enjoyed a renewed interest in the liquid-crystal community.
Their reduced dimensionality leads to peculiar properties of their phase transitions (nonexistence
of long-ranged order and defect-mediated phase transitions), but the  
possibility that these particles stabilise mesogenic (hence fluid) phases with high-order orientational 
symmetries (`exotic' nematic phases) has attracted much attention \cite{Review}. 
Schlacken et al. \cite{Schaklen} used the Scaled-Particle Theory (SPT) 
version of DFT, based on two-particle correlations via the second virial coefficient, to analyse
fluids of hard rectangles, and an intermediate tetratic phase, with four-fold symmetry, was predicted to
stabilise for relatively low aspect ratios. This is in contrast to the standard uniaxial nematic phase
with twofold symmetry in systems of rods with head-tail symmetry. Later, a more complete fluid phase diagram 
was obtained using the same theory \cite{M-R5}, and the effect of three-body correlations was assessed \cite{M-R3}
and seen to reinforce tetratic order. Other convex
hard particles with polygonal shape have been examined using
the same theoretical approach, and phases with triatic, or sixfold, symmetry, were predicted in fluids of
particles with isosceles triangular shape close to the equilateral condition \cite{M-R2}. More recently,
fluids of hard right-angled triangles (HRT)
have been seen to present evidences of strong octatic, or eightfold, orientational correlations, 
which constitutes a challenge for the standard DFT theories \cite{M-R4}.

On the simulation side, orientational order in fluids made of anisotropic hard particles has been 
studied by many authors. Nematic phases have quasi-long-range order and are obtained from the
isotropic through a continuous transition driven by a Kosterlitz-Thouless mechanism \cite{Frenkel0}.
Focussing on hard or quasi-hard particles of convex polygonal shape, phases
with tetratic phases were initially studied in squares \cite{Frenkel} and then confirmed
by several studies \cite{Donev,Krauth,Escobedo}. Hard rectangles have also been studied but with
not so much emphasis on the tetratic phase \cite{Anderson,Fitchhorn}. 
Recently fluids of HRT particles
have been seen to exhibit peculiar orientational properties \cite{Gantapara,M-R4}, with strong
indications of quasi-eightfold symmetry. Other particle shapes such as pentagons and
hexagons have been analysed \cite{Anderson}. As expected, the latter form a fluid 
phase with sixfold symmetry

Lithographic techniques are being applied in the production of particles with specific 
polygonal shapes \cite{Zhao0,Zhao1,Zhao2}. These particles behave as Brownian 
particles when in solution, and can be arranged in monolayers.
Also, filamentous virus particles can be engineered to control particle
stiffness. Ordering properties of such systems under confinement have been explored and compared
with elastic and microscopic theories \cite{Aarts}. These developments open up the possibility to 
experimentally study the effect of entropy on 2D orientational ordering and, in addition, to provide 
data against which standard and new theories can be contrasted. 

Another field of research in this area has emerged recently. Vertically vibrated monolayers of 
granular macroscopic particles confined into horizontal cavities
have been shown to provide a useful setting to probe 2D fluid ordering. 
Narayan et al. \cite{Narayan}
observed uniaxial nematic and tetratic phases in systems of quasi-2D metallic cylinders
(which project horizontally as quasirectangles).
Galanis et al. \cite{Galanis1,Galanis2} studied uniaxial nematic ordering of metallic
needles and concluded that the standard elastic theory of liquid crystals can be used
to explain some features of the experimental results, such as emergence or nematic order
and the competition with orientation at a surface boundary, in what appears to
be a kind of `entropic anchoring' mechanism \cite{Aarts0}. Using Monte
Carlo (MC) simulation, the equilibrium nematic and tetratic
phase behaviour of rectangles has been compared with the behaviour of vibrated monolayers
\cite{Dani}, and reasonable overall agreement was found in controlled regions of parameter space.
Also, comparison of clustering effects and ordering between the equilibrium and dissipative
systems, and also the formation of topological defects, has been made \cite{Miguel,Joe}. All these studies
point to a rich phenomenology and to subtle entropic effects giving rise to nontrivial
symmetries and response to external fields in systems of particles with relatively simple
shapes.

A striking effect in systems of regular, or not very nonregular, polygonal shapes, is the
occurrence of nematic phases with high symmetries, i.e. higher than the standard uniaxial symmetry
found in fluids of rods with head-tail particle symmetry. The tetratic phase formed by hard rectangles
of low aspect ratio is the simplest of these phases. In general polygonal-shaped particles can give rise to 
the so-called $p$-atic fluid phases. In these phases the orientational distribution function 
$h(\phi)$, giving the 
probability that, on average, a particle is oriented along some direction $\phi$ with respect to some fixed
axis, exhibits rotational symmetry by a multiple of $2\pi/p$, with 
$p=2, 3, 4\cdots$. The standard uniaxial nematic would have $p=2$.
Hard rectangular particles (including squares) have been shown to stabilise
into a tetratic phase, with $p=4$. Recently Anderson et al. \cite{Anderson} have conducted 
systematic simulations on various types of particles and shown that some regular polygonal 
particles exhibit $p$-atic phases:
equilateral triangles ($6$-atic), squares ($4$-atic), and hexagons ($6$-atic). 
Regular polygons with seven edges or more behave as hard discs in that the isotropic (I) phase
crystallises via first a continuous transition to a hexatic fluid and then a first-order transition
to a crystal, without liquid-crystal behaviour. However, polygons with fewer than seven edges
very easily align with respect to their neighbours and 
a local liquid-crystalline order is imposed, giving rise to
$p$-atic phases that transform into the isotropic via a KTHNY mechanism. Regular pentagons are
special because of the incompatibility of the corresponding $5$-atic symmetry with a regular
crystal and no liquid-crystal phase results, with an isotropic phase that directly crystallises
into a solid via a first-order phase transition.

However, there are strong indications that systems of particles made of {\it nonregular}
polygonal shapes 
may form intermediate orientational phases of nontrivial symmetries due to clustering mechanisms
\cite{M-R4}. In equilibrium the clustering tendency is driven by entropic forces 
\cite{entropic,entropic1}, and
additionally dissipation may contribute in vibrated monlayers \cite{Miguel}.
By nontrivial symmetry we mean a symmetry which does not arise directly from particle shape. 
In this work we discuss the case of HRT fluids, which seem to exhibit a $4$-atic phase but 
with very strong $8$-atic (or octatic) correlations. 
This symmetry can only result from the coexistence of several types of particle clusters, including
dimers and tetramers. Recent
simulation results \cite{M-R4} indicate that the
fractions of these clusters are such that the global phase is almost perfectly $8$-atic.
By contrast, the standard DFT approximation, based on the second and third
virial coefficients, is unable to deal with the high-order correlations
needed to explain the formation of some of these clusters. This is an
unusual situation in the theory of liquid crystals, as the standard DFT approach, 
originating from Onsager\cite{Onsager}, has been very successful to explain the occurrence and
symmetries of liquid-crystal phases both in 2D and 3D, at least at the qualitative level.
These negative results point to the need for a 
serious improvement of the theory in its application to 2D fluids of 
hard convex particles of nonregular polygonal shape.

In this article we describe some of our recent work on the orientational
phase behaviour of some 2D hard particles, which covers both
theoretical (predictions on phase diagrams based on DFT and MC simulation) 
and experimental results on monolayers of vibrated granular particles. The conclusions 
gravitate about three basic points: (i) The standard DFT, based on two-body
angular correlations, gives essentially correct predictions as to the
equilibrium orientational phases, provided the particles consist of regular
or close to regular polygons. (ii) For some non-regular polygonal particles
the tendency of particles to cluster into well-identified groups of
particles with a definite symmetry may be very strong; in these cases the
symmetry of the bulk phase may not be evident from the particle symmetry
and the DFT approach may fail dramatically. In particular, symmetries
that do not follow directly from particle shape, such as the $8$-atic
symmetry, may result from particle clustering
in the case of HRT. (iii) Vibrated monolayers
of particles show many common features with the equilibrium counterparts: bulk ordering and 
symmetry, response to external fields and excitation of defects under geometrical frustration.
These systems may represent a fruitful approach to study order in 2D
fluids of hard particles, complementary to experiments on Brownian particles and to
theoretical approaches.

The paper is organised as follows. In Section \ref{theory} we review the SPT version of DFT
theory applied to general 2D hard particles, and its extension to incorporate three-body
correlations through an orientational-dependent third-virial coefficient. Also, we show
some results for the $2$-atic and $4$-atic phases of hard rectangles (HR), and for the
$6$-atic phase of triangles (HT) close to equilateral, to make clear the connection between the bulk
symmetry and the (microscopic) orientational distribution function. Section \ref{theoretical_results}
is devoted to the theoretical results on phase diagrams. The fluid phase diagram of one-component
HR is discussed, and the strong effect of three-body correlations emphasised, 
pointing to the importance of clustering in the structure of these systems. 
Next, the phase behaviour
of a fluid of HT particles is shown, as a function of the opening angle and in the framework of
the SPT theory. In addition, the fluid of HRT is discussed, with a focus on the
evidence from simulation on the existence of strong eight-fold correlations in this system.
Both the standard SPT version of DFT, and the extended version with the third virial coefficient,
fail to reproduce this symmetry, without the slightest hint of secondary peaks in a completely
uniaxial orientational distribution function. We present evidence from simulation 
for the formation of strong clustering in this fluid, and discuss a model based on a quaternary
mixture of particles with shapes given by the clusters coexisting in the configurations 
observed in the simulations. This model, although not fully consistent, is an extension of SPT theory
and  provides a mechanism that can explain the presence of such strong $8$-atic correlations in the system.
In Section \ref{experiments} we present our results on experiments made on vibrated quasimonolayers
of metallic particles. Steady-state $4$-atic configurations are stabilised in monolayers of
cylinders (which behave as hard rectangles), and the imposed circular symmetry of the confining
cavity creates topological defects which agree in charge and structure with the predictions
of topology and equilibrium statistical mechanics. Other geometries are discussed and the
results explained in terms of the breakdown of the continuum hypothesis implicit in the 
elastic theory of liquid crystals. We end in Section \ref{conclusion} with some conclusions.

\section{Theory}
\label{theory}

In this section we review the standard DFT theory for 2D liquid crystals,
as applied to a fluid mixture of hard bodies. The theory is based on
an approximate expression for the excess or interaction 
part of the Helmholtz free-energy, $\beta {\cal F}_{\rm ex}[\{\rho_i\}]$, 
as a functional of the density profile of species $i$ ($i$ varying from 1 to the 
total number of components), $\rho_i({\bm r},\hat{\boldsymbol{\omega}})$, 
with ${\bm r}$ and $\hat{\boldsymbol{\omega}}$ the position vector of the centre of mass
and the unit vector 
parallel to the main particle axis of the $i$th species, respectively. We restrict our study 
to orientationally ordered uniform phases of 2D uniaxial particles, i.e. the 
density profiles, $\rho_i(\phi)$, only depend on a single angular coordinate, $\phi$, 
the angle between the main particle axis and one of the fixed Cartesian axis.

\begin{figure}
	\centering
		\includegraphics[width=3.in]{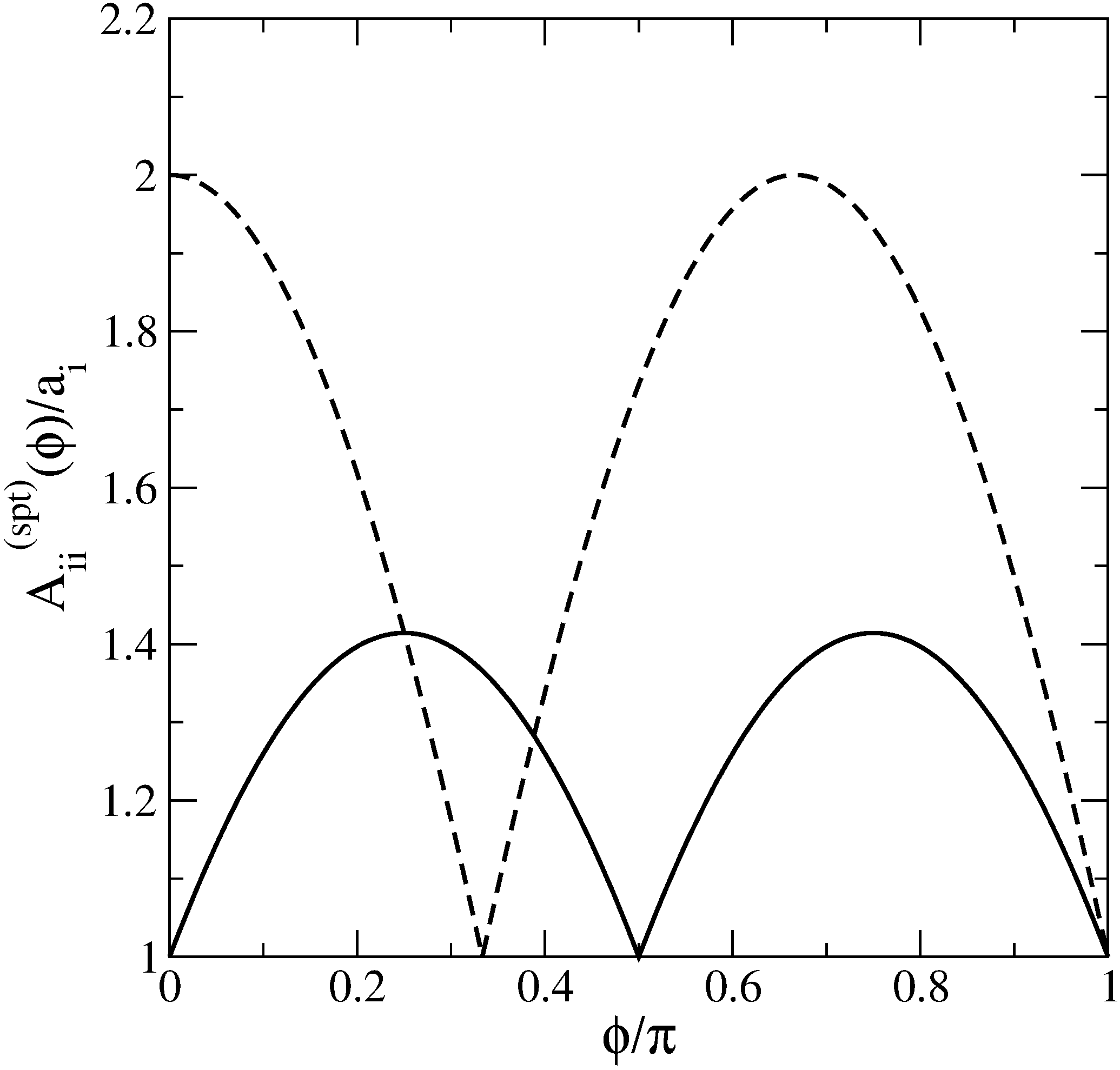}
	\caption{SPT areas (scaled with particle area) between two hard squares (solid) and two equilateral triangles 
	(dashed).}
	\label{fig1}
\end{figure}

One of the most commonly used versions of DFT is SPT
\cite{Reiss,Cotter1,Cotter2,Cotter3,Lasher,Barboy}. This theory successfully 
describes the phase behavior of hard-body fluids.
The 2D version of this theory for a general fluid mixture of
convex particles proposes the following expression for the excess part 
of the free-energy density:
\begin{eqnarray}
	\Phi_{\rm ex}[\{\rho_i\}]\equiv \frac{\beta {\cal F}_{\rm exc}[\{\rho_i\}]}{A}
	=-\rho \log(1-\eta)+\frac{\sum_{i,j}\langle\langle A^{(\rm spt)}_{ij}(\phi)\rangle\rangle}{1-\eta},
	\label{exceso}
\end{eqnarray}
with $\beta=(k_B T)^{-1}$ and $A$ the total area of the system. In the above expression 
the packing fraction of the mixture, $\eta$, is defined as 
\begin{eqnarray}
	\eta=\sum_i \rho_i a_i,\quad \rho_i=\int_0^{2\pi}  d\phi \rho_i(\phi),
\end{eqnarray}
with $\rho_i$ the number density of the $i$th species, while $a_i$ is its particle area. 
The total number density is just $\rho=\sum_i \rho_i$. The 
functions $A^{{\rm (spt)}}_{ij}(\phi)$ are defined in terms of the excluded area between
particles of species $i$ and $j$, with their axes forming a relative angle $\phi$:
\begin{eqnarray}
	A^{\rm (spt)}_{ij}(\phi)=\frac{1}{2}\left[A_{ij}^{(\rm excl)}(\phi)-a_i-a_j\right].
\end{eqnarray}
The symbol $\langle\langle \dots \rangle\rangle$ means the double angular average with 
respect to $\rho_i(\phi)$ and $\rho_j(\phi)$:
\begin{eqnarray}
	\langle\langle A^{\rm (spt)}_{ij}(\phi)\rangle\rangle\equiv 
	\int_0^{2\pi}d\phi_1\int_0^{2\pi}d\phi_2 \rho_i(\phi_1)\rho_j(\phi_2) 
	A^{\rm (spt)}_{ij}(\phi_1-\phi_2).
\end{eqnarray}
It is easy to show that the expression (\ref{exceso}) recovers the exact Onsager second-virial 
form in the low-density limit $\rho_i\to 0$. Also, more sophisticated versions of DFT, namely those 
based on the 
Fundamental Measure Theory, have SPT as their uniform density limit \cite{Tarazona}. 

\begin{figure}
	\begin{center}
		\includegraphics[width=6.in]{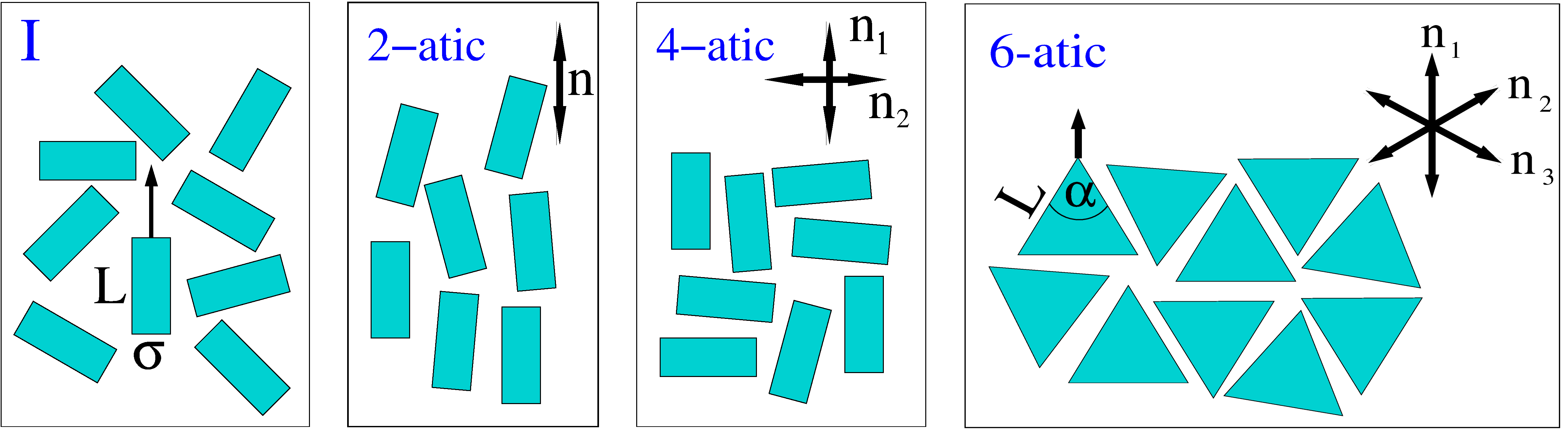}
	\caption{Sketches of liquid-crystal phases of HR and HT. Particle axes are indicated, together with the nematic directors of each phase.}
	\label{fig1_new}
	\end{center}
\end{figure}

\begin{figure}
	\begin{center}
	\includegraphics[width=3.in]{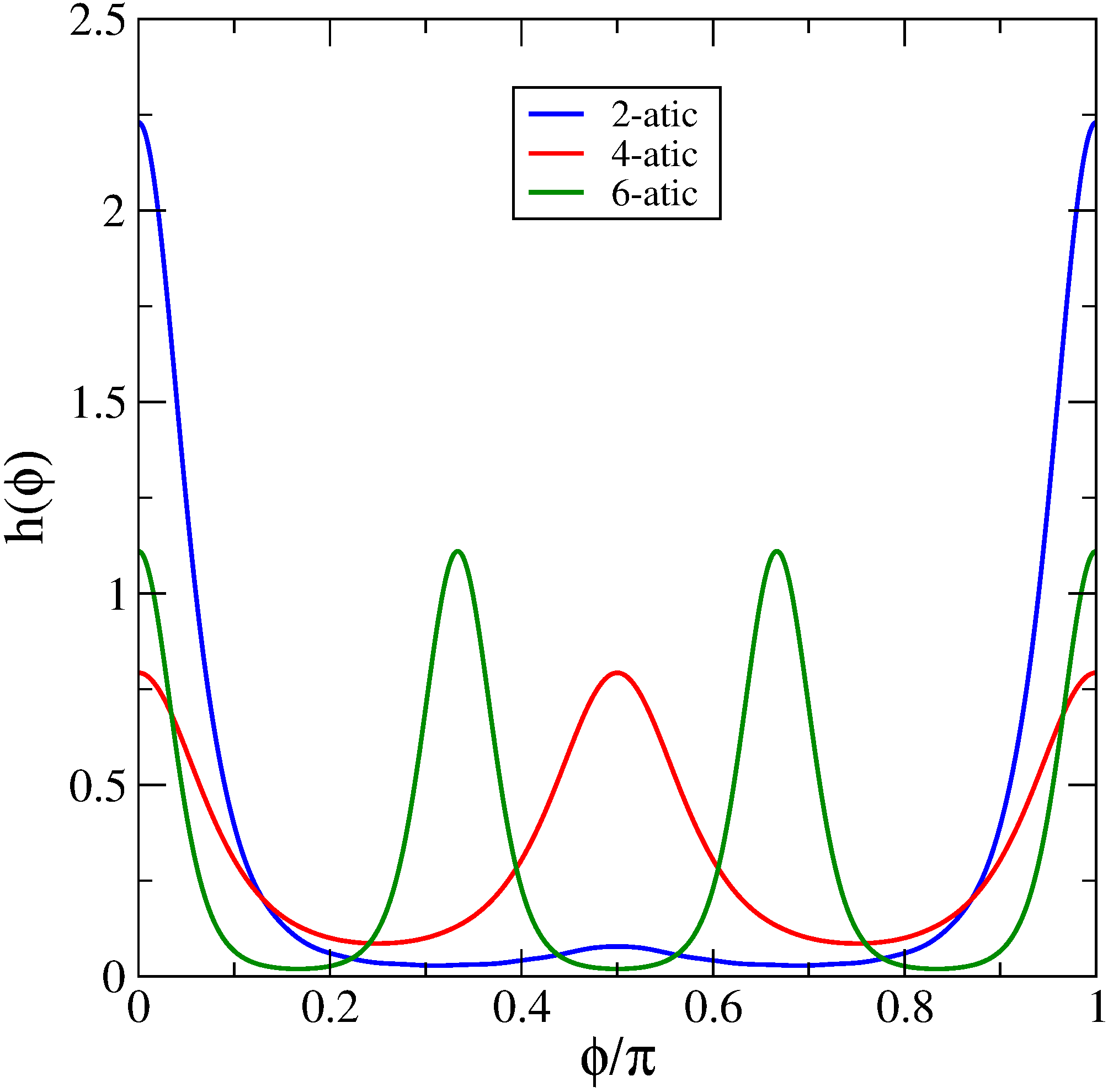}
		\caption{The typical orientational distribution functions $h(\phi)$ corresponding to 2-atic, 4-atic phases of HR, and 6-atic phase of HT.}
	\label{fig1_new2}
	\end{center}
\end{figure}

	In Fig. \ref{fig1} we plot two examples of SPT areas (scaled with particle area) 
	between two hard squares and two hard 
equilateral triangles. In these cases particles are chosen as having equal shape and 
size (same species).
We note the symmetry of these functions: for squares the excluded area has a $\pi/2$-symmetry, 
$A_{11}^{(\rm spt)}(\phi)=A_{11}^{(\rm spt)}(\phi+\pi/2)$, while for equilateral 
triangles a $2\pi/3$-symmetry, $A_{22}^{(\rm spt)}(\phi)=A_{22}^{(\rm spt)}(\phi+2\pi/3)$, 
is apparent.
These symmetries in turn imply that the only possible orientationally ordered phases 
for squares and triangles must have orientational invariance by angles of $\pi/2$ and 
$\pi/3$, i.e. $\rho_1(\phi)=\rho_1(\phi+\pi/2)$ and 
$\rho_2(\phi)=\rho_2(\phi+\pi/3)$, respectively. These are the so called
$4$-atic and $6$-atic orientationally ordered phases. 
Note that the $2\pi/3$-symmetry of the excluded area 
of equilateral triangles, together with the absence of a polar phase 
(the nematic directors and their $\pi$-rotated counterparts are equivalent) implies 
the $\pi/3$ symmetry of triangles.
In Fig. \ref{fig1_new} we show sketches of the I, 2-atic and 4-atic phases for one-component HR and the 6-atic phase of triangles, while their typical probability density orientational distribution functions are shown in Fig. \ref{fig1_new2}. These functions 
are defined in terms of the density profile $\rho(\phi)$ as 
$h(\phi)\equiv \rho(\phi)/\rho$ (with $\int_0^{2\pi} d\phi h(\phi)=1$). Note the presence of one (2-atic), two (4-atic) and three (6-atic) peaks of the same height (in $[0,\pi]$) of the function $h(\phi)$, which shows the two-fold (2-atic), fourfold (4-atic) and sixfold (6-atic) orientational liquid-crystal symmetries.

The ideal part of the free-energy density is exact and has the expression 
\begin{eqnarray}
	\Phi_{\rm id}[\{\rho_i\}]\equiv \frac{\beta {\cal F}_{\rm id}[\{\rho_i\}]}
	{A}=\sum_i \int_0^{2\pi} \rho_i(\phi)\left[ \ln \rho_i(\phi)-1\right],
\end{eqnarray}
where we have dropped the thermal areas.

The total free-energy density, 
$\Phi[\{\rho_i\}]=\Phi_{\rm id}[\{\rho_i\}]+\Phi_{\rm exc}[\{\rho_i\}]$, should be minimized 
with respect to $\rho_i(\phi)$ to find the equilibrium configuration, for  
fixed values of the number densities $\rho_i$. The constrained minimization and the inversion of the 
minimized ideal part gives the following expression for $\rho_i(\phi)$:
\begin{eqnarray}
	&&\rho_i(\phi)=\frac{\rho_ie^{-c_i(\phi)}}{\int_0^{2\pi} d\phi' e^{-c_i(\phi')}},
	\\ 
	&& c_i(\phi)=\frac{\delta \Phi_{\rm exc}[\{\rho_i\}]}{\delta \rho_i(\phi)}
	=-\ln(1-\eta)+\tilde{c}_i(\phi) +\beta p a_i, \\ 
	&&\tilde{c}_i(\phi)=\frac{2\sum_j \int_0^{2\pi} d\phi' \rho_j(\phi') 
	A^{(\rm spt)}_{ij}(\phi-\phi')}{1-\eta},\\
	&&\beta p=\frac{\rho}{1-\eta}+\frac{\sum_{i,j}\langle\langle A_{ij}^{(\rm spt)}(\phi)\rangle\rangle}
	{(1-\eta)^2} \label{pressure},
\end{eqnarray}
As the terms $-\ln(1-\eta)$ and $\beta p a_i$ (the scaled pressure) are the same in the
exponentials of the numerator and denominator and do not depend on the angular coordinate, they cancel out, which results in
\begin{eqnarray}
	\rho_i(\phi)=\frac{\rho_ie^{-\tilde{c}_i(\phi)}}{\int_0^{2\pi} d\phi' e^{-\tilde{c}_i(\phi')}}
	\label{inversion}
\end{eqnarray}
We now perform a bifurcation analysis to find the packing fraction value at which the I 
phase becomes unstable with respect to the 2-atic or 4-atic phases. 
It is useful to use a Fourier expansion of the density profiles: 
\begin{eqnarray}
	\rho_i(\phi)=\rho_i h_i(\phi)=\frac{\rho_i}{2\pi}\left[1+\sum_{k\geq 1} h_i^{(k)}
	\cos(2k\phi)\right], 
	\label{la_rho}
\end{eqnarray}
with $h_i(\phi)$ the orientational distribution function of species $i$, and $h_i^{(k)}$ its 
$k$th Fourier amplitude. We define the Fourier coefficients of the SPT area as
\begin{eqnarray}
	A_{ij,k}^{(\rm spt)}\equiv \frac{1}{2\pi}\int_0^{2\pi} d\phi A_{ij}^{(\rm spt)}(\phi)
	\cos(2k\phi).
\end{eqnarray}
The analytic form of these coefficients can be found for hard-rectangle mixtures and  
hard-isosceles-triangle mixtures in Ref. \cite{M-R1} and \cite{M-R2}, respectively.
With these definitions, inserting Eqn. (\ref{la_rho}) into Eqn. (\ref{inversion}), 
multiplying by $\cos(2n\phi)$, integrating from $0$ to $2\pi$, and expanding the 
exponentials up to first order in the first Fourier amplitudes $h_i^{(n)}$ (with
$n$ denoting the symmetry, and ($n=1$ for $2$-atic, $n=2$ for $4$-atic, and $n=3$ 
for $6$-atic) we finally obtain a set of algebraic equations,
\begin{eqnarray}
	B\cdot {\bm h}^{(n)}={\bm 0}, \ 
	B_{ij}=\delta_{ij}+\frac{2A_{ij,n}^{(\rm spt)}\rho_j}{1-\eta},
\end{eqnarray}
with ${\bm h}^{(n)}$ the column vector formed by the Fourier amplitudes $\{h_i^{(n)}\}$ of all species from $i=1$ up to 
$i=c$ (the total number of components).
This set has a nontrivial solution only if ${\cal B}(\eta)\equiv\text{det}\left (B\right)=0$. The 
latter is just the condition to find the packing fraction at bifurcation. For example for 
a binary mixture we find 
\begin{eqnarray}
	{\cal B}(\eta)=1+2\frac{\rho \overline{A^{(\rm spt)}_{ii,n}} }{1-\eta}
	+\frac{4\rho_1\rho_2}{(1-\eta)^2}\text{det}\left( A_n^{(\rm spt)}\right),
\end{eqnarray}
where we have defined 
\begin{eqnarray}
	\overline{A^{(\rm spt)}_{ii,n}}=\sum_i x_i A_{ii,n}^{(\rm spt)},\quad 
	\text{det}\left( A_n^{(\rm spt)}\right)=A_{11,n}^{(\rm spt)}A_{22,n}^{(\rm spt)}-
	\left(A_{12,n}^{(\rm spt)}\right)^2,
\end{eqnarray}
with $x_i=\rho_i/\rho$ the molar fraction of the $i$th species. Taking into account that,
for the particle geometries we are considering in this work, we have 
$\text{det}\left( A_n^{(\rm spt)}\right)=0$, from the bifurcation condition ${\cal B}(\eta)=0$ we obtain the packing fraction at bifurcation as 
\begin{eqnarray}
	\eta_n=\frac{1}{1-2 \overline{A^{(\rm spt)}_{ii,n}}/\overline{a_i}},
\quad \overline{a_i}=\sum_i x_i a_i.
	\label{bifurca}
\end{eqnarray}
The degree of orientational ordering of species $i$ in the mixture can be quantified through the 
order parameters $Q_2^{(i)}$ ($2$-atic) and $Q_4^{(i)}$ ($4$-atic) for HR,
and also $Q_8^{(i)}$ ($8$-atic) for HRT 
and $Q_6^{(i)}$ ($6$-atic) for equilateral triangles. They are defined as 
\begin{eqnarray}
	Q_{2n}^{(i)}=\int_0^{2\pi}d\phi h_i(\phi) \cos(2n\phi)=\frac{h_i^{(n)}}{2}.
	\label{order_parameters}
\end{eqnarray}
In the case where a mixture exhibits a first-order transition, the two-phase 
coexistence curves are calculated through the equalities of the chemical potentials of the
coexisting phases 
$\alpha$ and $\beta$ for each species, and the equality of pressure at $\alpha$ and $\beta$,
both of which guarantee chemical and mechanical equilibrium, i.e.
\begin{eqnarray}
	\mu_i(\{\rho_i^{(\alpha)}\})=\mu_i(\{\rho_i^{(\beta)}\}),\quad 
	p(\{\rho_i^{(\alpha)}\})=p(\{\rho_i^{(\beta)}\}),
\end{eqnarray}
where the chemical potentials and pressure are calculated from $\Phi[\{\rho_i\}]$ as 
\begin{eqnarray}
	\beta \mu_i=\frac{\partial \Phi}{\partial \rho_i}, \quad 
	\beta p=\sum_i \rho_i (\beta \mu_i)-\Phi. 
\end{eqnarray}
The explicit expression for $\beta p$ is given by Eqn. (\ref{pressure}), while the chemical potential of species $i$, 
$\mu_i=\mu_i^{(\rm id)}+\mu_i^{(\rm exc)}$, can be divided into ideal and excess part. The ideal part is 
\begin{eqnarray}
	\beta \mu_i^{(\rm id)}=\ln \rho_i+\int_0^{2\pi} d\phi h_i(\phi) \ln h_i(\phi),
\end{eqnarray}
while the excess part can be computed as 
\begin{eqnarray}
	\beta \mu_i^{(\rm exc)} =-\ln(1-\eta)+\frac{2\sum_j \rho_j\langle\langle A_{ij}^{(\rm spt)}(\phi)\rangle\rangle_{h_{i,j}}}{1-\eta}+\beta p a_i,
\end{eqnarray}
where
\begin{eqnarray}
\langle\langle A_{ij}^{(\rm spt)}(\phi)\rangle\rangle_{h_{i,j}}\equiv
	\int_0^{2\pi}d\phi h_i(\phi)\int_0^{2\pi}d\phi' h_j(\phi')
	A_{ij}^{(\rm spt)}(\phi-\phi').
\end{eqnarray}
For a one-component fluid all the above expressions are the same but sums over species and
indexes are omitted.
The extension of SPT to include not only the second, $B_2[h]$, but also the third, $B_3[h]$, virial coefficient 
for the one-component fluid can be implemented 
by replacing the expression (\ref{exceso}) for the excess part of the free-energy density by
\begin{eqnarray}
	\Phi_{\rm exc}[\rho]=\rho\left\{-\ln(1-\eta)+\frac{\eta}{1-\eta}b_2[h]+\left(\frac{\eta}{1-\eta}+
	\log(1-\eta)\right)\left(b_3[h]-2b_2[h]\right)\right\},
	\label{B3}
\end{eqnarray}
where we have defined the scaled (with particle area $a$) virial coefficients as 
\begin{eqnarray}
	b_k[h]=\frac{B_k[h]}{a^{k-1}}-1.
	\end{eqnarray}
The coefficient $b_2[h]$ is just the doubled angular average of the scaled  
function $A^{(\rm spt)}(\phi)$ with respect to the orientational distribution function $h(\phi)$, i.e. $b_2[h]=a^{-1}\langle\langle A^{(\rm spt)}(\phi)\rangle\rangle_{h}$.
	The third virial coefficient is defined as 
	\begin{eqnarray}
		&&B_k[h]=\frac{1}{3}\prod_{i=1}^3 \left(\int_0^{2\pi} d\phi_i h(\phi_i)\right) {\cal K}^{(3)}
		(\phi_1,\phi_2,\phi_3),\\
		&&{\cal K}^{(3)}(\phi_1,\phi_2,\phi_3)=-\frac{1}{A} \left(\prod_{i=1}^3 \int_A d{\bm r}_i\right)
		f({\bm r}_{12},\phi_{12})f({\bm r}_{23},\phi_{23})f({\bm r}_{13},\phi_{13}),
	\end{eqnarray}
	with $f({\bm r}_{ij},\phi_{ij})$ the Mayer function of particle $i$ and $j$, with relative 
	spatial and angular variables ${\bm r}_{ij}={\bm r}_i-{\bm r}_j$ and $\phi_{ij}=\phi_i-\phi_j$ 
	respectively. For details of the bifurcation analysis using this 
	extended theory see Refs. \cite{M-R3,M-R4}.

\section{Theoretical results}
\label{theoretical_results}
This section presents the results obtained from the numerical implementation of the SPT 
for rectangular and triangular particles of one-component fluids. The case of  
a quaternary mixture of HRT is also discussed as a model for a fluid with strong 
clustering effects. We concentrate on the 
description of phase diagrams and bifurcations to orientationally ordered phases. Special 
attention is paid to the effect of clustering on the orientational properties of HRT.

\subsection{One-component fluids}
\label{one-component_fluids}

\begin{figure}
	\begin{center}
	\includegraphics[width=3.in]{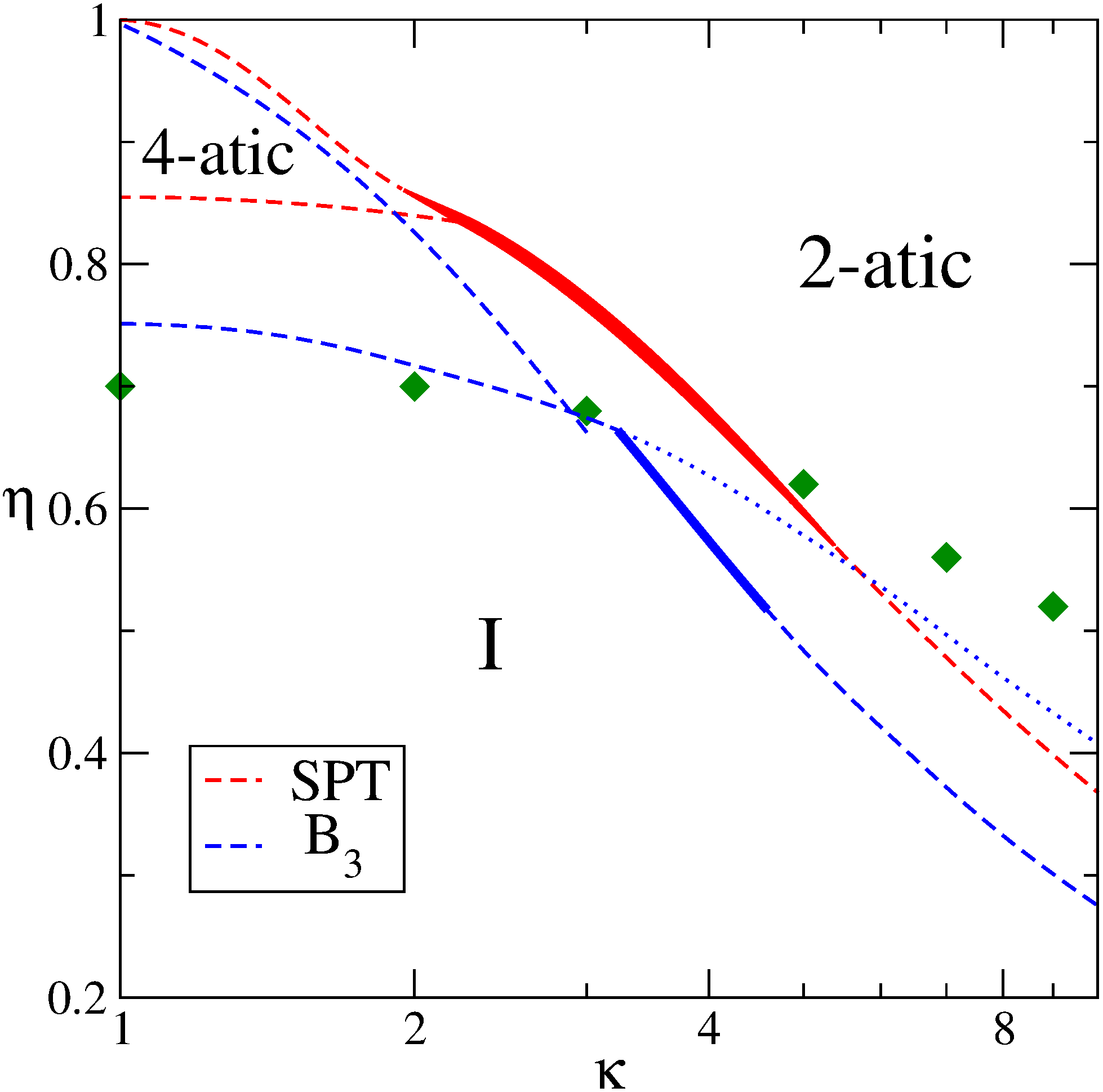}
		\caption{Phase diagrams of HR in the $\eta$-$\kappa$ plane from the
		standard SPT approximation and the extension including $B_3$ (as 
		indicated in the legend). Solid and dashed lines indicate first- and 
		second-order transitions, with filled regions corresponding to two-phase coexistence 
		gaps. Labels correspond to regions of stability of I, $4$-atic and $2$-atic phases.
		The dotted line is the extension of the I-T bifurcation curves from the $B_3$-theory 
		to the $\kappa$ values where $4$-atic is unstable with respect to the $2$-atic phase.
		Our MC simulation results for the location of the I-T (for $\kappa\lesssim 7$)
		or I-N transition are shown with filled rhombuses.}
	\label{fig2}
	\end{center}
\end{figure}

We define a 2D liquid-crystal model of particles consisting of monodisperse HR of length  $L$ and width $\sigma$, 
with aspect ratio $\kappa=L/\sigma\geq 1$. The SPT area between two rectangles with relative angle $\phi$ between their main axes 
(the main axis being parallel to the long axis $L$), which is the main ingredient of SPT, 
can be calculated as
\begin{eqnarray}
	A^{(\rm spt)}(\phi)=\frac{L^2+\sigma^2}{2}|\sin \phi|+L\sigma|\cos \phi|.
\end{eqnarray}
As shown in Sec. \ref{theory}, this expression defines the theoretical model completely. 
Note that, since the fluid is monodisperse, 
subindices for the different species can be eliminated.

MC simulations of hard squares (HS) \cite{Frenkel}
and HR of aspect ratio $\kappa=2$ \cite{Donev}
showed the existence of a $4$-atic stability window between the I and the crystalline phase. 
The latter, in the case of HR of $\kappa=2$, also exhibits $4$-atic ordering with square 
dimers at perpendicular orientations, tesseling the whole area at close packing.
Recent simulations showed that the $4$-atic phase is stable up to $\kappa\sim 5$ 
\cite{Taylor}, while experiments on monolayers of granular cylinders of aspect ratios as large as 7 
showed the existence of stationary states exhibiting 4-atic textures \cite{Narayan,Dani}.

The first calculation of the I--$4$-atic and I--$2$-atic second-order transitions for a HR 
fluid was done in Ref. \cite{Schaklen}, which established the 
stability of the 4-atic phase for $\kappa<2.61$ \cite{Schaklen}. However the two-phase 
coexistence lines and the $2$-atic--$4$-atic second-order transition were not calculated. The numerical implementation of SPT for the HR fluid gives the phase diagram plotted 
in Fig. \ref{fig2}, in the packing fraction ($\eta$)-aspect 
ratio ($\kappa$) plane \cite{M-R5}. We only accounted for uniform phases, but here
the coexistence binodals and the $2$-atic--$4$-atic spinodal line are also included.
It can be seen that, at high $\eta$, SPT predicts 
a stable 4-atic phase after a second-order I--$4$-atic transition for $\kappa\lesssim 
\kappa_{\rm ce}\approx 2.21$. This is the position of the critical-end point, where 
the I--$4$-atic bifurcation curve (calculated from Eqn. (\ref{bifurca}) with $n=2$) 
and the (I,$4$-atic)-binodal of the first order (I,$4$-atic)--$2$-atic transition coalesce.
The 4-atic phase stability region is bounded above by a $2$-atic--$4$-atic 
transition, which can be of first or second order, with the $2$-atic--$4$-atic 
tricritical point located at $\kappa_{\rm t_1}\approx 1.94$. For 
$\kappa>\kappa_{\rm ce}$, the I phase exhibits a first-order transition to the 2-atic 
phase up to $\kappa\leq \kappa_{\rm t_2}\approx 5.44$, the position of the I--$2$-atic 
tricritical point. Beyond this value of aspect ratio, the I--$2$-atic transition is 
always of second order and the bifurcation cirve is given by Eqn. (\ref{bifurca}) with $n=1$. 
The results from the $B_3$ theory are also plotted in 
Fig. \ref{fig2} \cite{M-R3}. The main differences with respect to the SPT are: (i) 
The I--$4$-atic bifurcation 
occurs at much lower packing fractions. (ii) The position of the critical end-point moves to 
higher values, $\kappa_{\rm ce}\approx 3.1$. These two facts together cause 
the stability region of the 4-atic phase to be much wider. Also, the I--$2$-atic 
transition curve moves to lower packing fractions when three-body correlations are included. 
The onset of stability for the 4-atic phase compares favourably with our own MC simulations
(rhombuses for $\kappa\lesssim 7$). This result points 
to the relevance of three-body, and possibly higher-order, correlations to quantitatively 
predict the phase behavior of 2D fluids of hard anisotropic particles.

\begin{figure}
	\begin{center}
		\includegraphics[width=3.in]{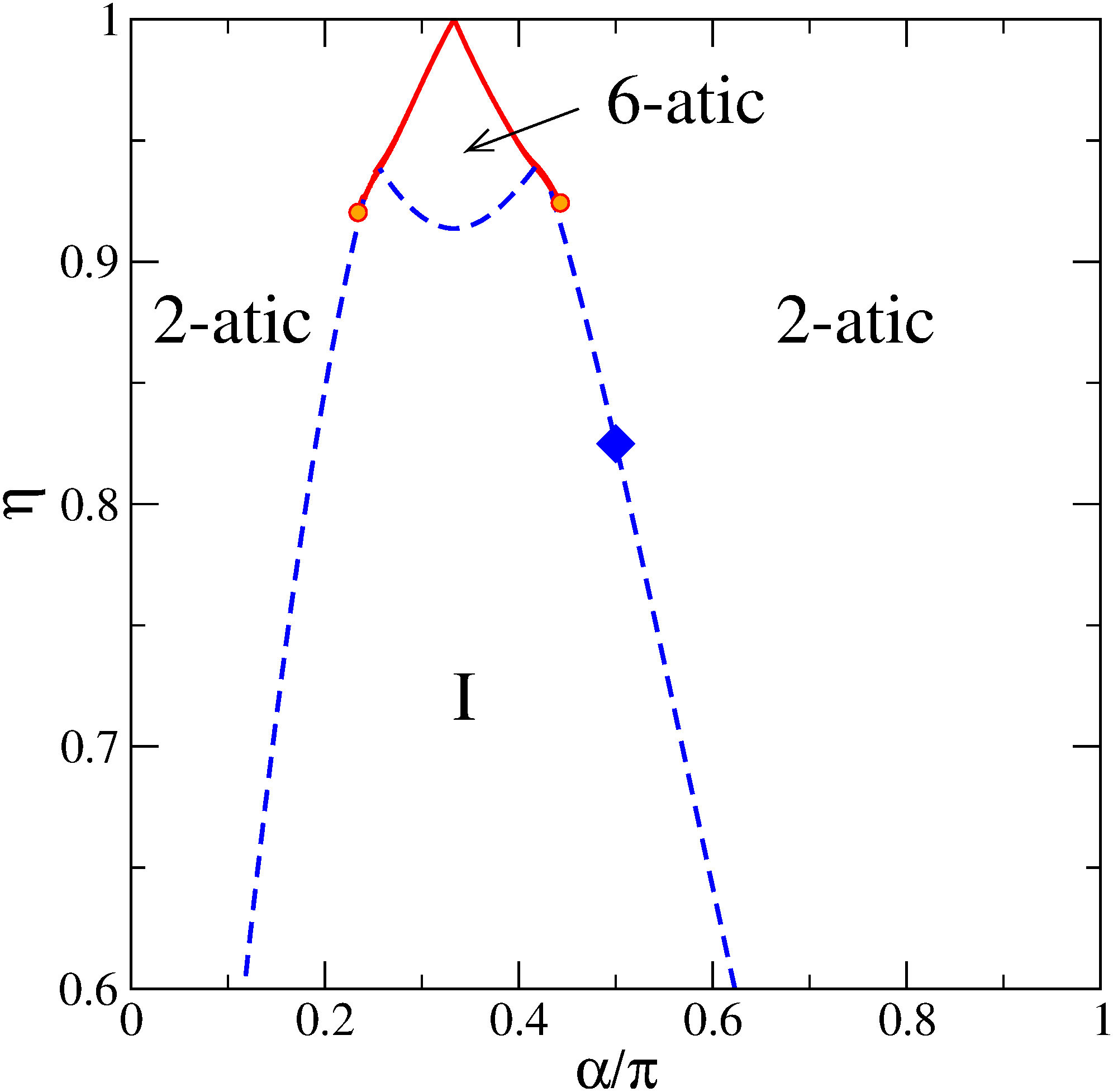}
		\caption{Phase diagram of HT in the $\eta$-$\alpha$ plane from
		SPT. With dashed lines 
		we show the second-order transition lines while the solid lines indicate 
		first order transitions. The regions of stability of I and the orientationally ordered 
		2-atic and 6-atic phases are labeled. With solid circles we show I$-$2-atic critical points. 
		The rhombus indicates the packing fraction value of the 
		I$-$2-atic bifurcation of HRT ($\alpha=\pi/2$).}
	\label{fig3}
	\end{center}
\end{figure}
We now turn to triangular particles.
A collection of isosceles triangles of equally sized edge-lengths $L$ and  
opening angle $\alpha$ define a one-component fluid of hard isosceles triangles. The SPT area 
between two particles with relative orientation $\phi$ (the particle axis defined to be parallel 
to the bisector of the opening angle) can be calculated from 
\begin{eqnarray}
	A^{(\rm spt)}(\phi)=L^2\left\{
		\begin{matrix}
			\sin \alpha \cos \phi, & 0\leq \phi\leq \alpha,\\
			\frac{1}{2}\sin(\alpha+\phi), & \alpha\leq \phi\leq (\pi-\alpha)/2,\\
			\sin\phi-\frac{1}{2}\sin(\alpha+\phi), & 
			(\pi-\alpha)/2\leq \phi\leq \pi.
		\end{matrix}
		\right.
		\end{eqnarray}
		for $\alpha\leq \pi/3$, while for $\alpha\geq \pi/3$
\begin{eqnarray}
	A^{(\rm spt)}(\phi)=L^2\left\{
		\begin{matrix}
			\sin \alpha \cos \phi, & 0\leq \phi\leq (\pi-\alpha)/2,\\
			(1-\cos \alpha)\sin\phi, & (\pi-\alpha)/2\leq \phi\leq \alpha,\\
			\sin\phi-\frac{1}{2}\sin(\alpha+\phi), & 
			\alpha\leq \phi\leq \pi.
		\end{matrix}
		\right.
		\end{eqnarray}
		In SPT these expressions completely define the model and the thermodynamics of the fluid. The phase diagram in the packing fraction $\eta$-opening angle $\alpha$ plane 
		is plotted in Fig. \ref{fig3}. Equilateral triangles, due to their intrinsic sixfold symmetry (invariance with respect to $\pi/3$-rotation), can only form an orientationally ordered 6-atic phase (see Fig. \ref{fig1_new}). But isosceles triangles 
		with opening angle $\alpha\simeq\frac{\pi}{3}$ 
		also stabilize this phase via a second order I$-$6-atic transition, shown 
		as a dashed line in Fig. \ref{fig3}. The 6-atic phase is bounded above by a first-order $6$-atic--$2$-atic transition, which ends in two critical points (filled circles).
		As the opening angle is increased up to $\pi$, or decreased down to 0,
		obtuse or acute triangles turn into hard needles. 
It is clear that if the aspect ratio of the particles, defined through the opening angle 
		as $\kappa=\cot(\alpha/2)$, increases or decreases, the fluid exhibits 
		a second order I$-$2-atic at high enough 
		packing fractions (dashed line of Fig. \ref{fig3} separating the regions of I and 2-atic phase stability). Although it is reasonable to expect 
		that particles with $\kappa\ll 1$ or $\kappa\gg 1$ (close to the Onsager limit) will stabilize the 2-atic phase, the situation 
		is not clear for isosceles triangles with opening angle around $\pi/2$, 
		i.e. hard right triangles (HRT), whose location in the phase diagram of
		Fig. \ref{fig3} is shown by a rhombus.

		Recent simulations \cite{Gantapara,M-R4} showed that a fluid of HRT stabilizes a 4-atic phase, with the presence of 8-atic correlations, and that the 
		orientational distribution function, $h(\phi)$, has a total of 
		eight peaks in the interval $[0,2\pi]$. If all the peaks had exactly the 
		same height the global phase would be 8-atic. As can be seen from the figure, 
		there is no indication of 4-atic or 8-atic phase stability for $\alpha\sim \pi/2$.
		Motivated by the fact that this result could be 
		a defect of the SPT (a second-virial theory), we have implemented
		the I to $p$-atic bifurcation analysis (with $p=2,4,6$ and $8$), using 
		the $B_3$ theory, with an excess  
		free-energy density given by Eqn. (\ref{B3}) \cite{M-R4}. 
		The results are shown in Table \ref{table}. Clearly the lowest value of
		packing fraction corresponds to the I$-$2-atic bifurcation. 
		A free-energy minimization for values of $\eta$ above this bifurcation 
		confirms the 2-atic phase stability, with the 
		equilibrium distribution function showing the usual peaks at 0 and $\pi$ 
		(see Fig. \ref{fig1_new2}) and the absence of any satellite peaks.
		The reason why the DFT with two or three-body 
		correlations incorporated cannot predict the 4-atic or 8-atic phase stability will be explained in Sec. \ref{mixtures}.

\begin{table}
\begin{tabular}{lcccc} \toprule
Bifurcation & I--$2$-atic  & I--$4$-atic & I--$6$-atic & I--$8$-atic \\ 
 $\eta_n$ from SPT & 0.8249 & 0.9928 & 0.9821 & 0.9444\\
 $\eta_n$ from $B_3$-theory & 0.7325 & 0.9794 & 0.9328 & 0.8353 \\
 \toprule
\end{tabular}
	\caption{Values of the packing fractions $\eta_n$ at I--2-atic ($n=1$), I--4-atic ($n=2$), 
	I--6-atic ($n=3$) and I--8-atic ($n=4$) bifurcations from the SPT and the $B_3$-theory.} 
\label{table}
\end{table}

\subsection{Clustering effects in fluids of triangles modeled as a quaternary mixture}
\label{mixtures}

\begin{figure}
	\begin{center}
	\includegraphics[width=3.5in]{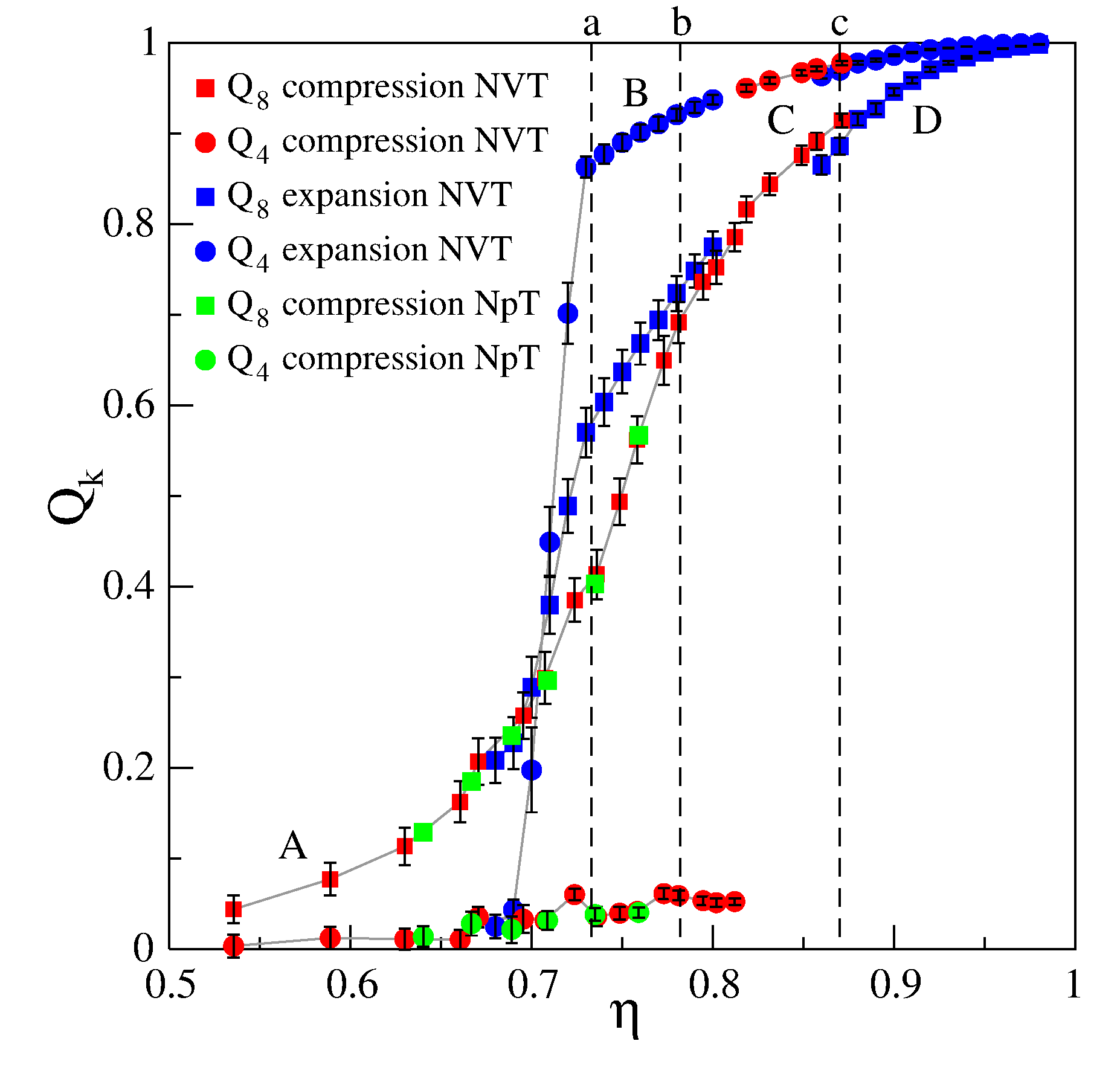}
		\caption{Order parameters $Q_4$ and $Q_8$ vs. packing fraction $\eta$ obtained 
		from NVT and NPT MC simulations of a fluid of HRT. Different expansion and compression 
		protocols were taken, as indicated. Label A indicates the region of stability of the I 
		phase, label B the I--$4$-atic phase coexistence region, and labels
	C and D indicate the region of stability 
		of 4-atic and crystal phases, respectively, as they were obtained in Ref. \cite{Gantapara}.}
	\label{fig4}
	\end{center}
\end{figure}

To explore the symmetry and stability of fluids made of HRT particles,
we have performed NVT and NPT MC simulations of systems of 576 particles \cite{M-R4}. 
Different compression and expansion protocols were performed. Fig. \ref{fig4} shows
the 4-atic and 8-atic order parameters $Q_4$ and $Q_8$ as a function of packing fraction $\eta$. 
Clearly, compression of the fluid from the I phase brings about an increase in the $Q_8$ order 
parameter, while the 4-atic order parameter $Q_4$ stays at a low value. This points to
strong 8-atic correlations in the fluid, with an orientational distribution function (not shown) 
having eight peaks of similar height in the interval $[0,2\pi]$. When compression is continued up 
to high packing fractions, the system remains in the 8-atic phase. Expanding a previously prepared
$4$-atic crystal (with a square unit cell made of four self-assembled triangles forming a tetramer)
from very high packing fractions, a stable crystal is obtained down to the melting point,
where a 4-atic liquid-crystal phase is formed with high values of $Q_4$ and $Q_8$ 
vanishing small (the nature of this phase transition cannot be ascertained from our results). 
This 4-atic phase remains stable up to $\eta\sim 0.72$, below which $Q_4$ abruptly drops to zero. 
This behavior of $Q_{2n}$ shows that the system exhibits strong hysteresis in the region where 
the 8-atic and 4-atic phases are present, an indication that the I$-$4-atic transition is strongly
of first order. The MC free-energy calculations of the 4-atic and crystalline phases carried out 
by the authors of Ref. \cite{Gantapara} indicate that the 4-atic phase is the stable liquid-crystal 
phase. In Fig. \ref{fig4} label B indicates the two-phase coexistence region 
obtained from Ref. \cite{Gantapara}, while regions C and D correspond to the regions of stability 
of the 4-atic and crystal phases, respectively \cite{Gantapara}. 

We have also prepared a crystalline phase with a unit cell now made of 
square dimers obtained by joining two triangles by their hypotenuses, and 
the hypotenuses of all triangles 
being parallel to each other along the crystal. This is a perfect crystal with $Q_2=1$. After expanding this 
crystal, it eventually melts 
to a 2-atic liquid-crystal phase with a high value of $Q_2$ \cite{M-R4}. 
The system continues to be metastable up to values of $\eta$ when 
$Q_2$ drops to zero. This behavior implies that, although the 2-atic phase is metastable, the free-energy difference between 
the 2-atic and 4-atic phases should be small. 

\begin{figure}
	\begin{center}
	\includegraphics[width=5.in]{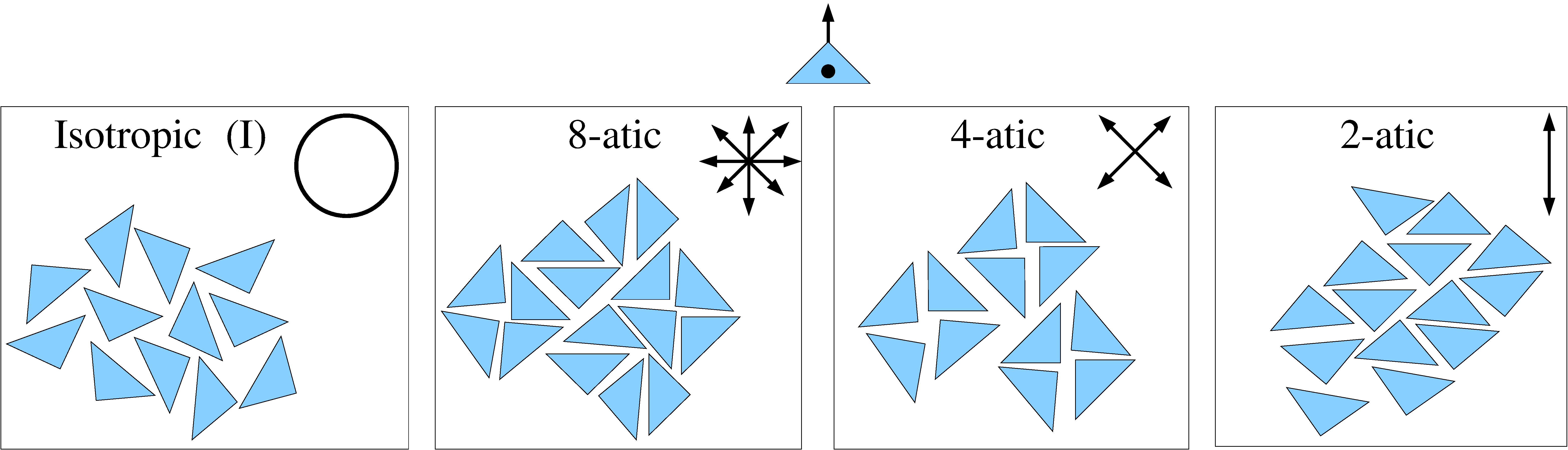}
		\caption{Sketches of I, 8-atic, 4-atic and 2-atic phases of a HRT fluid, where 
		strong clustering of 
		triangles 
		can induce the corresponding liquid-crystal symmetries. Also shown are the equivalent 
		directors of these phases, having twofold (2-atic), fourfold (4-atic) and eightfold (8-atic) 
		symmetries.}
		\label{fig5}
	\end{center}
\end{figure}

From the preceding discussion we can conclude that the fluid of HRT exhibits strong 8-atic correlations by 
expanding the I phase from low densities, but the 4-atic is the stable liquid-crystal phase. 
As pointed out in Sec. \ref{theory} both, 
the SPT and the $B_3$-theory (the latter including two and three-body correlations), 
predict that the I phase bifurcates to a 2-atic phase, which remains stable above the bifurcation. 
Thus the DFT fails to correctly predict the symmetry of the stable liquid-crystal phase. This failure  
is related to the fact that the standard DFT formalism, based on the one-body density, cannot take into account 
the effect of clustering or self-assembling of particles. If we try to assemble a 4-atic or 8-atic (apart from I and $2$-atic) 
phase made of HRT by resorting to configurations with strong particle clustering,
something similar to the configuration sketched in Fig. \ref{fig5} results. 
Note how a 4-atic phase can be obtained by the presence of 
a large amount of square tetramers, formed by self-assembling of four triangles 
(the monomer units) joined by their short edge-sides. The diagonals of these square 
tetramers are oriented on average along two mutual 
perpendicular and equivalent directions. The 8-atic phase could be obtained by the presence of these tetramers, but a large amount of square dimers formed by two triangles joined by their hypotenuses have to be added. If these hypotenuses are oriented 
along the same directions of the tetramer orientations, the fluid will exhibit 
strong 8-atic correlations, i.e. the orientational distribution function will have eight 
sharp peaks in the interval $[0,2\pi]$.

\begin{figure}
	\begin{center}
	\includegraphics[width=2.5in]{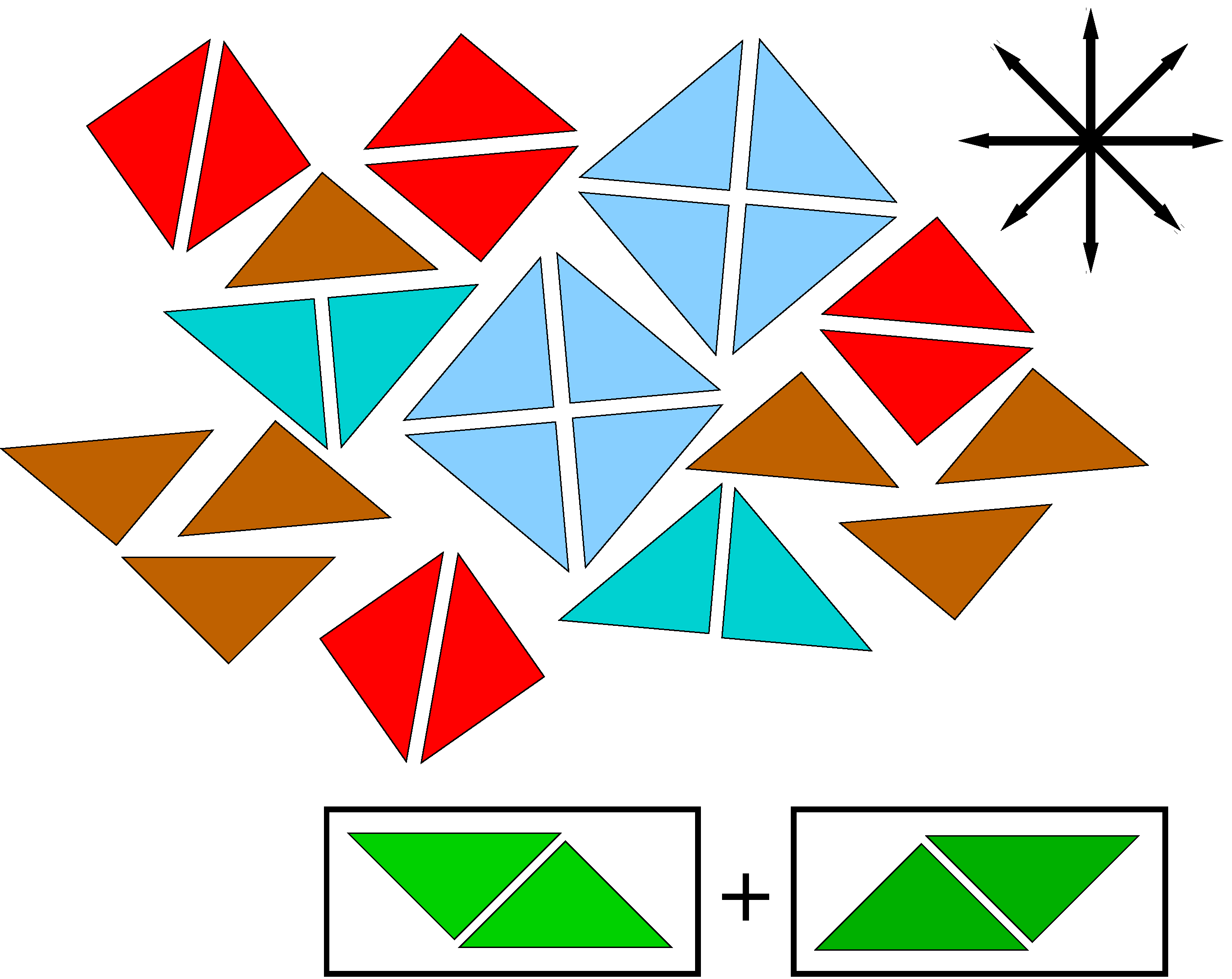}
		\caption{Sketches of the four types of clusters used in the theoretical model,
		represented in different colors. 
		The rhomboidal clusters discarded in our model are shown at the bottom.}
	\label{fig6}
	\end{center}
\end{figure}

A very simplified model to account for particle clustering is to treat the one-component 
fluid of HRT as a quaternary mixture, with species defined as following (see Fig. \ref{fig6}): (i) The first species is of square geometry and made by a dimer formed by joining two triangular monomer units by
their hypotenuses. This species will have a molar fraction $x_1^{(1)}$ 
(we adopt the convention of using the 
subindex to label the geometry: 1 for the square and 2 for the triangle, and the superindex to label the 
size: 1 for small and 2 for large size), and has $L_1^{(1)}=L$ as the square side length and 
$a_1^{(1)}=L^2$ as particle area (see Table \ref{tab}). (ii) The second species again is square but 
defined by a tetrameral cluster, i.e. four triangles joined by their short sides, with a molar fraction 
$x_1^{(2)}$, side length $L_1^{(2)}=\sqrt{2}L$ and area $a_1^{(2)}=2L^2$. (iii) The third species 
is a triangular monomer unit of side length
$L_2^{(1)}=L$ and area $a_2^{(1)}=L^2/2$, with the molar fraction labelled as $x_2^{(1)}$. And 
(iv) the last species is a triangular dimer formed by two monomer units joined by their short sides,
with the right angles of the triangles 
being in contact (see Fig. \ref{fig6}); the equally sized side lengths are $L_2^{(2)}=\sqrt{2}L$,
the area $a_2^{(2)}=L^2$, and the molar fraction $x_2^{(2)}$. Obviously 
these triangular species are needed in our model because two of them, with the adequate relative 
orientation, will form a tetramer. See Table \ref{tab} for a summary of the parameters. Note 
that there are more possible dimers: one formed by selecting one of the two possible configurations 
resulting (i) by sticking two triangular units by their short sides, but with the right and 
acute angles of the triangles in contact, and (ii) the other from a specular reflection of the 
latter with respect to the axis passing through the triangular heights (see Fig. \ref{fig6}). 
We discarded these dimers with rhomboidal shape for the sake of simplicity but,
as shown later, they certainly appear in MC simulations. 

We used the SPT described in Sec. \ref{theory} to numerically minimize the total free-energy with respect to the 
 orientational distribution functions of all four species $\{h_i^{(j)}(\phi)\}$. The aim
 was to look for the existence of a set of molar 
 fractions $\{x_i^{(j)}\}$ for which the resulting distribution function of monomers $h_{\rm m}(\phi)$ has 
 8-atic symmetry. The sum over species is taken over the subindexes and superindexes of 
 all quantities involved in the SPT. For example the SPT area now reads 
 $A_{ij}^{({\rm spt},kl)}(\phi)$, and depends on the set of the characteristic side lengths of species 
 $\{L_i^{(j)}\}$, and the sum over species of the double angular average should be 
 substituted by $\sum_{i,j,k,l=1,2} \langle\langle A_{ij}^{({\rm spt},kl)}(\phi)\rangle\rangle$. The 
 total packing fraction of the mixture is now defined as $\eta=\sum_{i,j=1,2} \rho_i^{(j)} a_i^{(j)}$ with 
 $\rho_i^{(j)}=x_i^{(j)}\rho$ the number density of species $ij$. 

\begin{figure}
	\begin{center}
		\includegraphics[width=2.2in]{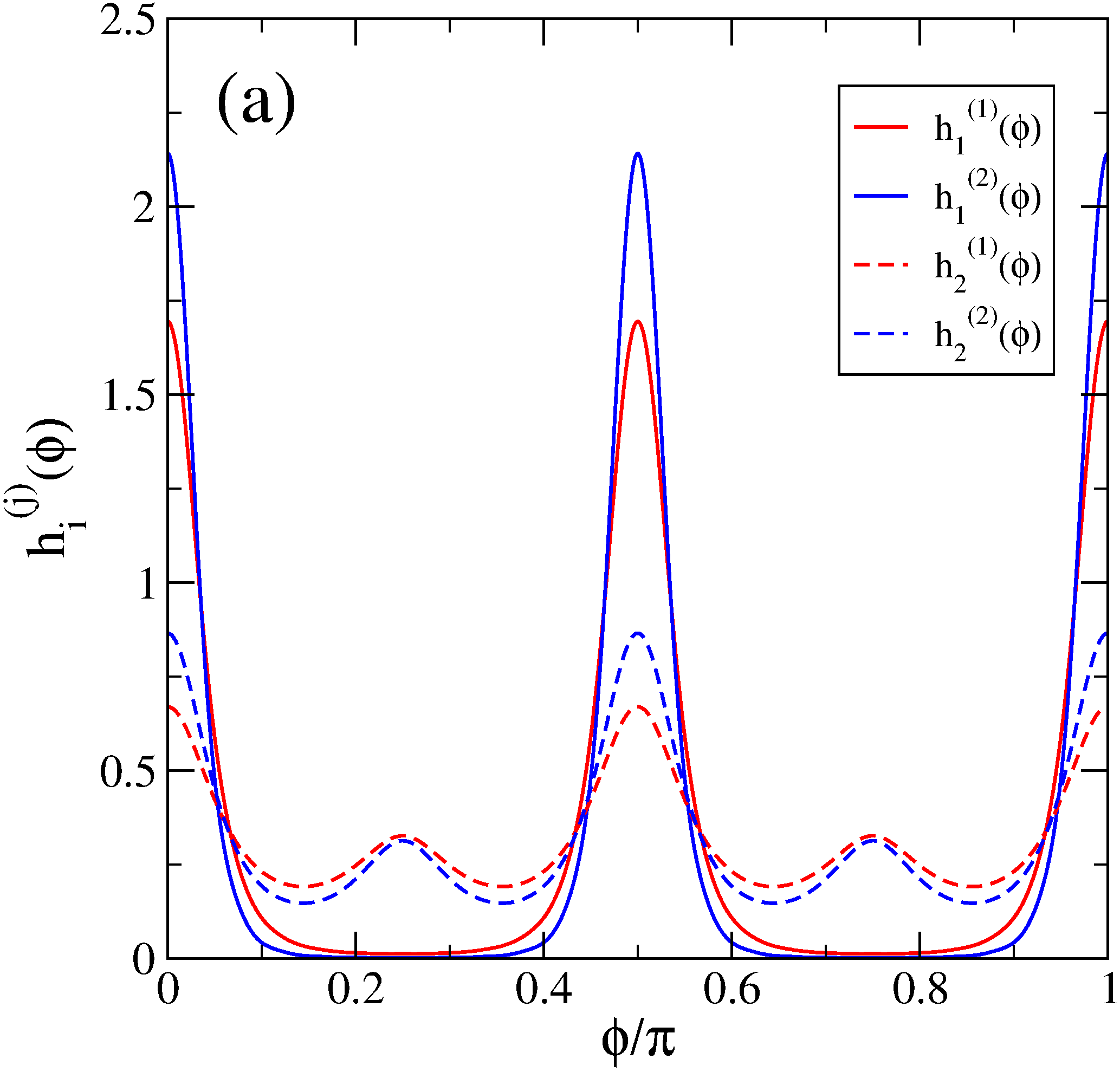}
		\includegraphics[width=2.2in]{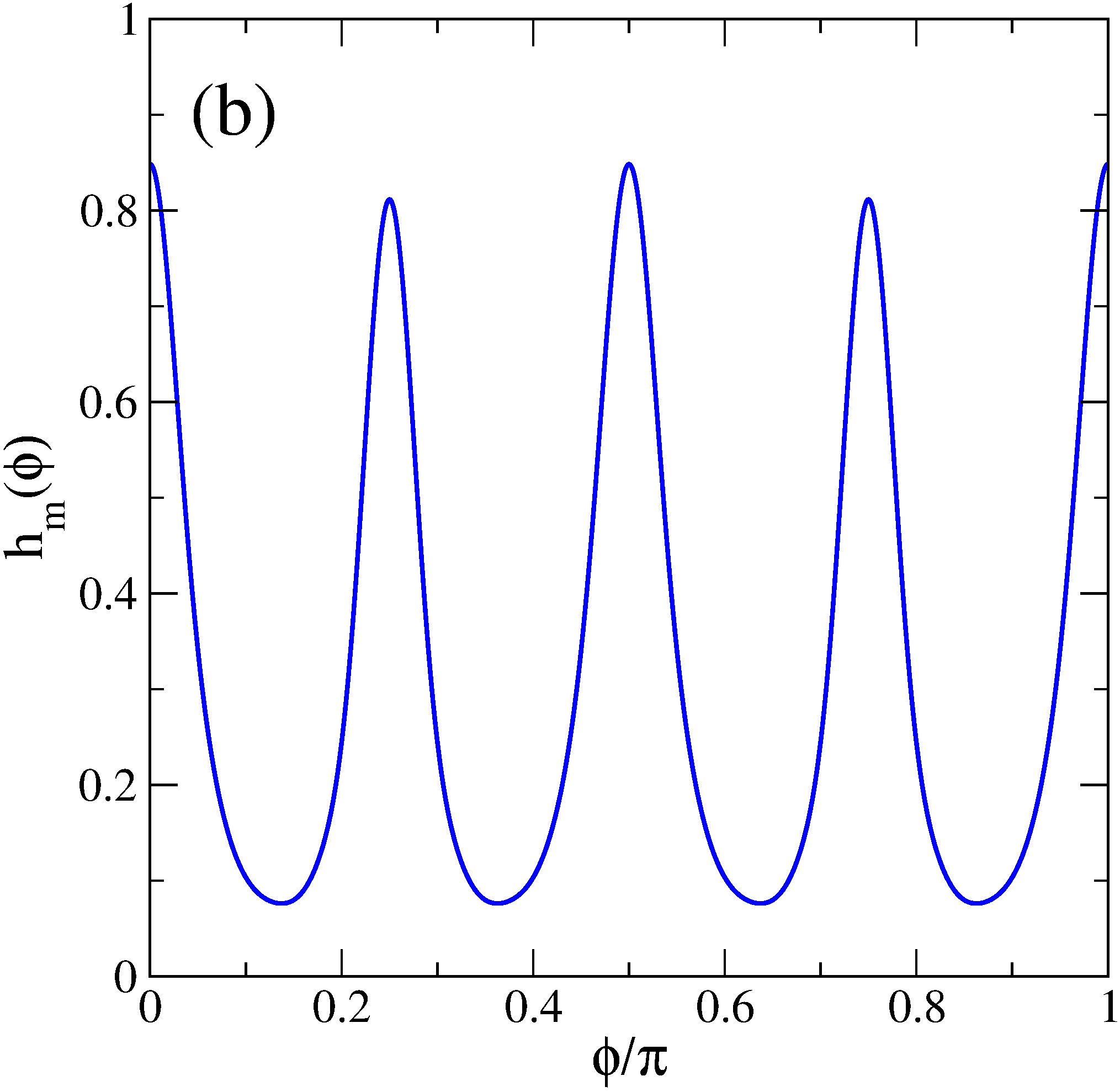}
		\includegraphics[width=2.2in]{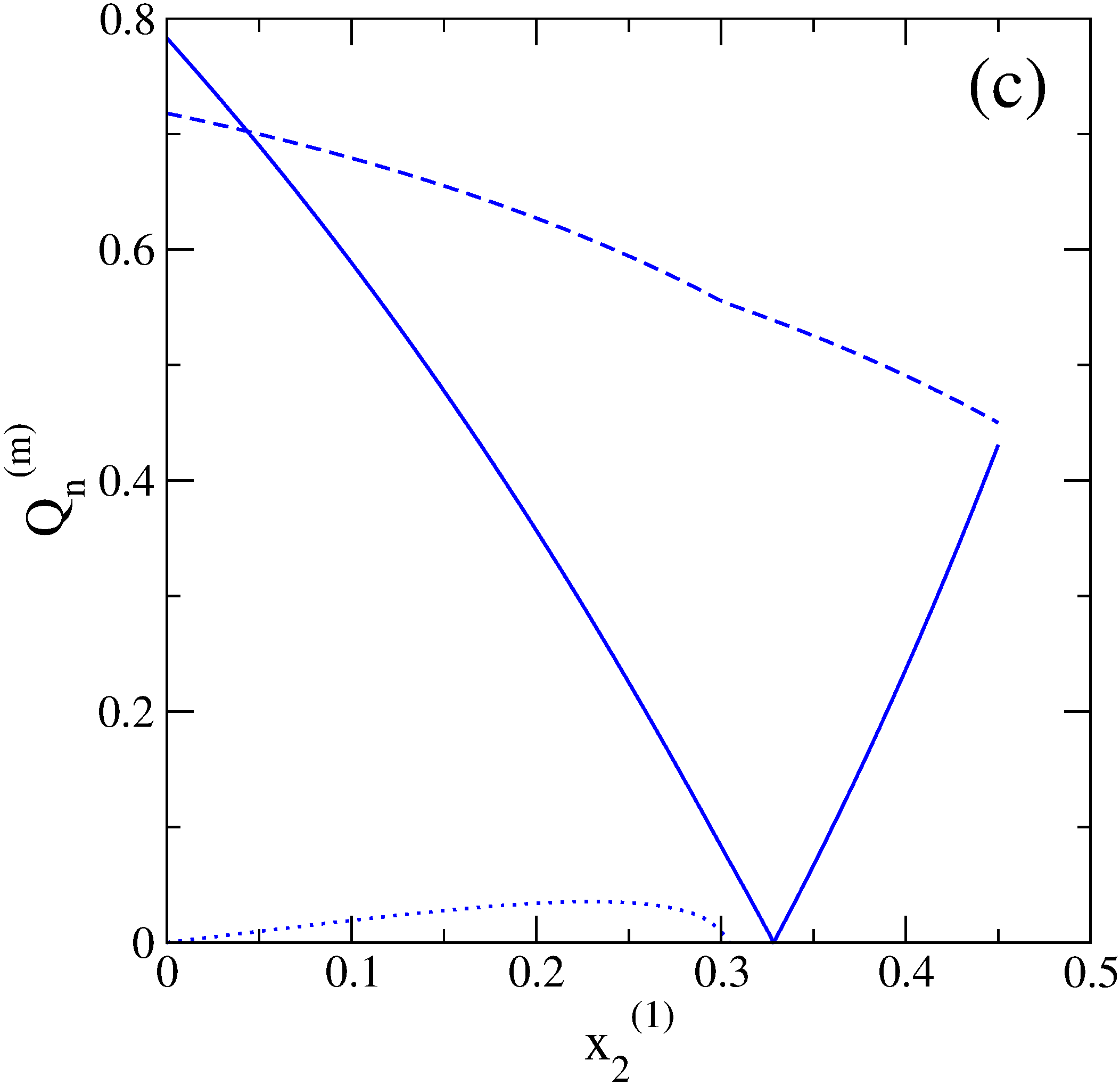}
	\caption{(a) Orientational distribution functions of all four species for the set of molar fractions 
	$x_1^{(1)}=0.4$, $x_1^{(2)}=0.15$, $x_2^{(1)}=0.35$, and $x_2^{(2)}=0.1$. Scaled pressure 
	if fixed to $\beta p a_2^{(1)}=220$. (b) The resulting monomer distribution function $h_{\rm m}(\phi)$. 
	(c) Order parameters $Q_2$ (dotted), $Q_4$ (solid) and $Q_8$ (dashed) as a function of the 
	molar fraction of monomers $x_2^{(1)}$ with the constraint 
	$x_1^{(1)}=x_2^{(1)}+0.05$, $x_1^{(2)}=0.5-x_2^{(1)}$, $x_2^{(2)}=0.45-x_2^{(1)}$. Note that these 
	constraints are such that $x_1^{(1)}+x_2^{(2)}=x_1^{(2)}+x_2^{(1)}$.}
	\label{fig7}
	\end{center}
\end{figure}

 From $\{h_i^{(j)}(\phi)\}$ we can calculate the monomer distribution function $h_{\rm m}(\phi)$, i.e. the 
 probability density of triangular monomer units being oriented with respect to a fixed laboratory axis with 
 angle $\phi$. The expression 
 that relates these magnitudes can be found in Ref. \cite{MR7}. From $h_{\rm m}(\phi)$ we can calculate the 
 order parameters as $Q_{2n}^{(m)}=\int_0^{\pi}d\phi h_{\rm m}(\phi) \cos(2n\phi)$. In Fig. \ref{fig7} (a) we show all 
 the distribution functions $\{h_i^{(j)}(\phi)\}$ for the following fixed values of molar fractions: 
 $x_1^{(1)}=0.4$, $x_1^{(2)}=0.15$, $x_2^{(1)}=0.35$ and $x_2^{(2)}=0.1$. The scaled pressure was fixed to 
 $\beta p a_2^{(1)}=220$. As we can see, these distribution functions correspond to a 4-atic phase, with all distribution 
functions fulfilling the fourfold symmetry $h_i^{(j)}(\phi)=h_i^{(j)}(\phi+\pi/2)$. This phase is the equilibrium phase, due to the presence of a large amount of square clusters, whose fourfold symmetry dictates the symmetry of the phase.
 Interestingly, triangular monomers and dimers follow the same 4-atic symmetry of square clusters. 
However their distribution functions exhibit the presence of two more satellite peaks of lower height
at $\phi=\pi/4$ and $\phi=3\pi/4$. Although $h_2^{(1,2)}(\phi)$ does not fulfill the $8$-atic symmetry 
(invariance with respect to rotations by $\pi/4$), triangular clusters certainly exhibit some 8-atic 
correlations, which is directly related to the crossed excluded area between squares and triangles (the 
expression can be found in Ref. \cite{MR7}). This function has local minima at relative angles of
$\pi/4$ and $3\pi/4$. But the most remarkable result 
is the symmetry of the angular distribution function of monomers $h_{\rm m}(\phi)$ resulting from the
$\{h_i^{(j)}(\phi)\}$ functions; this is shown in Fig. \ref{fig7} (b). 
We can see the quasi-$8$-atic symmetry, $h_{\rm m}(\phi)
\sim h_{\rm m}(\phi+\pi/4)$, of monomers. This result proves that strong clustering of particles into 
superparticles of square and triangular symmetries can induce 8-atic ordering, the one observed 
in MC simulations by compressing the I fluid. In panel (c) 
the order parameters of monomers $Q_{2n}^{(\rm m)}$ are plotted as a function of the monomer molar 
fraction 
$x_2^{(1)}$ for the same fixed pressure and along the following path of molar fractions: 
$x_1^{(1)}=x_2^{(1)}+0.05$, $x_1^{(2)}=0.5-x_2^{(1)}, x_2^{(2)}=0.45-x_2^{(1)}$ with $x_2^{(1)}$ 
varying in the interval $[0,0.45]$. There is an interval about $x_2^{(1)}\approx 0.33$ 
where the order parameter $Q_8$ is the highest one as compared to $Q_2$ and $Q_4$. This
indicates the presence of strong 8-atic correlations.
Strong 8-atic correlations are also obtained for other selected paths (not shown here), and it can
be concluded that, for some cluster compositions, a quasi$-$8-atic symmetry can be easily obtained. 
Of course this model is oversimplified in the sense that the molar fractions are fixed. An extended
SPT theory, supplemented by chemical equilibrium conditions between 
different clusters and monomers at fixed packing fraction of monomers, together with the addition 
of rhomboidal clusters, would allow to find the equilibrium values of $x_i^{(j)}$ and also explore
the effect of rhomboidal clusters on the symmetry of liquid-crystal 4-atic and 8-atic phases.

\begin{figure}
	\begin{center}
		\includegraphics[width=3.5in]{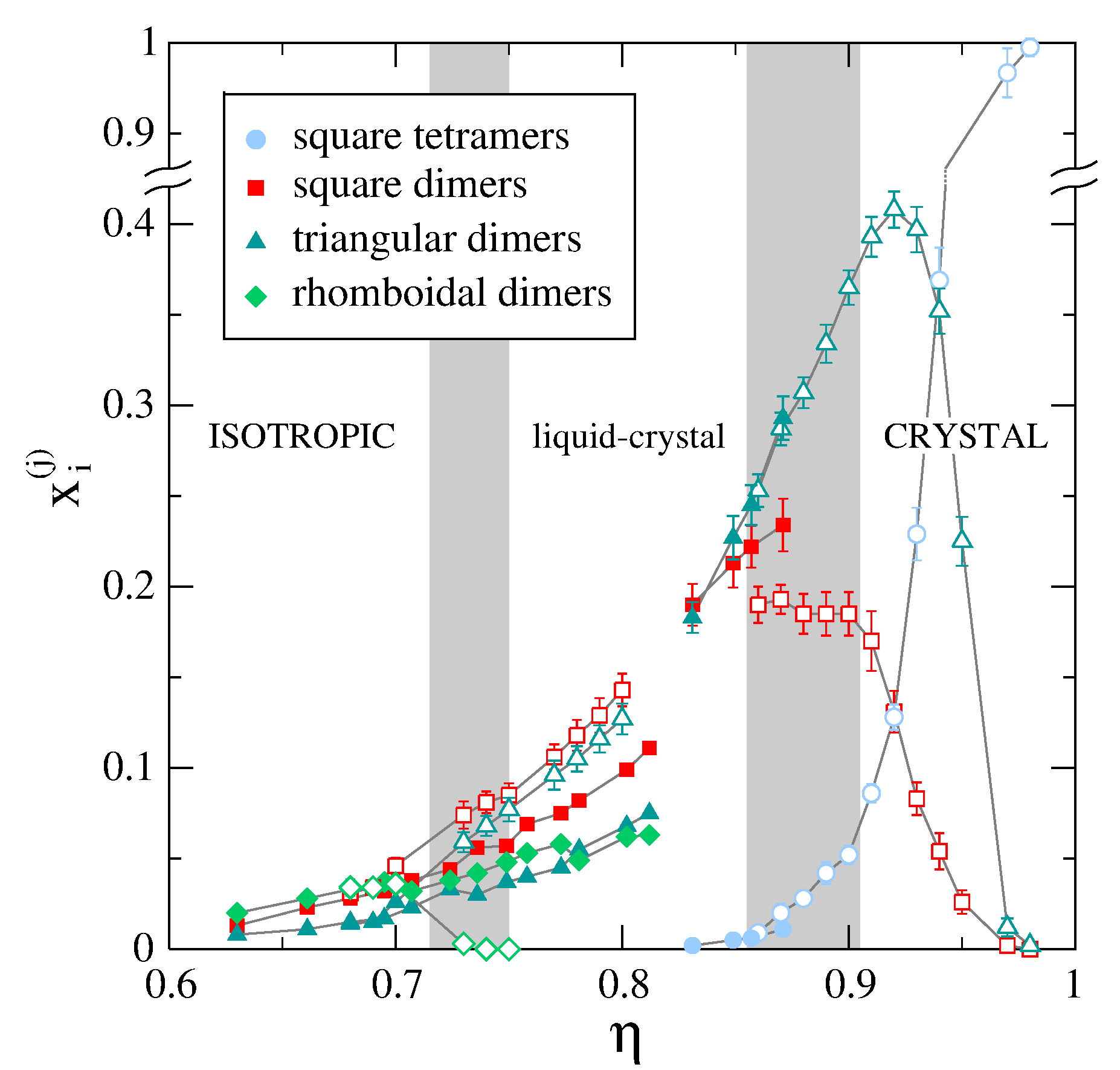}
		\caption{ 
		Cluster compositions $x_i^{(j)}$ as a function of $\eta$ as obtained from MC simulations 
		using compression (filled symbols) and expansion (open symbols) runs. See Ref. \cite{MR7} 
		for details. The regions of stability of the different phases are correspondingly labelled.}
		\label{fig8}
	\end{center}
\end{figure}

Fig. \ref{fig8} shows the results of a clustering analysis from 
equilibrated MC simulations of HRT at different packing fractions. 
Details of this analysis can be found in Ref. \cite{MR7}. Two important features are apparent.
when the I fluid is compressed (filled symbols) all clusters, except the monomer units, are more or 
less equally represented: see the packing-fraction region between 0.7 and 0.82 inside the liquid-crystal 
stability region. The compression run from the I fluid gave rise to 8-atic correlations.
However, expansion runs from the 4-atic crystal (open symbols) at high packing fraction show
strong clusterisation, with an excess of tetramers and the subsequent appearance of triangular and 
square dimers (but no rhomboidal dimers) which exhibit maxima as the packing fraction decreases and
the crystal melts to the 4-atic liquid-crystal phase. Both set of branches 
(from low and high densities) approximately connect to each other.
 
\begin{table}
	\centering 
	\epsfig{file=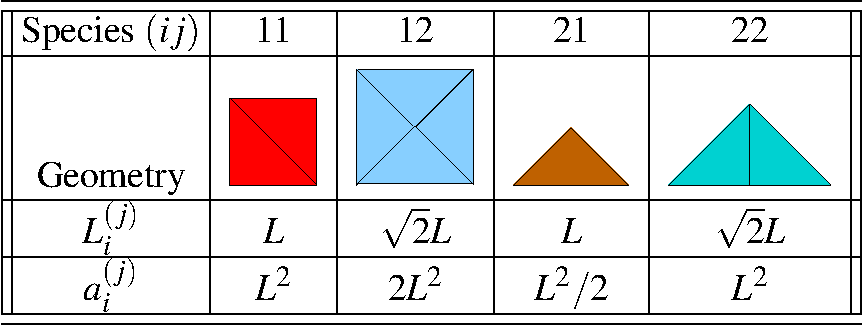,width=3.in}
	\caption{Summary of the geometries of all four species that define the clustering model.
		Also shown are the values of 
		side lengths $L_i^{(j)}$ and particle areas $a_i^{(j)}$ of the different species.}
	\label{tab}
\end{table}

As shown in this section, the standard DFT is very successful in explaining
the symmetry of the equilibrium phases in the case of fluids made of hard rectangles and 
also of HT. However, in the case of HR, the extension of the theory to include the third
virial coefficient points to the importance of high-order correlations and clustering.
These effects are not enough to modify the symmetry of the bulk phases. But the HRT fluid
is an example where these effects may be crucial, to the extent that the stable bulk
phases may have a symmetry which does not follow from the particle shape, or from the symmetry
of the excluded area between two particles.

\section{Experiments on vibrated granular particles}
\label{experiments}

Vibrated monolayers of granular particles may be a valid approach to explore the symmetries
of hard particles in 2D and the effects of entropy on the orientational properties of
these fluids. In this sense they could play a role complementary to theory and simulation.
Pioneering work by Narayan et al. \cite{Narayan} on vertically vibrated monolayers of anisotropic 
particles showed that granular particles such as rice, metallic cylinders or metallic pinwheels,
can organise into liquid-crystalline 2D configurations. These configurations exhibit qualitatively
similar properties as their thermal counterparts. For instance, in the steady-state regime,
well-defined values of various order parameters can be obtained. Galanis et al. \cite{Galanis1,Galanis2}
have also shown that some aspects of ordering in granular rods, when confined into cavities, 
can be explained in terms of the standard elastic free energy of nematics in competition with
a surface energy. Since these early works, some effort has been devoted to map
the phase diagram of cylinders \cite{Dani,Miguel}, which project as quasi-rectangles, as a function
of packing fraction, based on the values of $2$-atic and $4$-atic order parameters.
These studies showed that phase boundaries between isotropic, $2$-atic $4$-atic
regions reasonably coincide with those derived from equilibrium DFT calculations.
More recently, we have shown \cite{Miguel,Joe} that geometrical frustation induced by circular boundaries 
gives rise to the excitation of topological defects in monolayers of cylindrical particles, which 
seem to obey the same rules, based on topological charge conservation, as in equilibrium systems 
governed by a continuous tensor order parameter.
\begin{figure}[h]
        \begin{center}
	\includegraphics[width=4.5in]{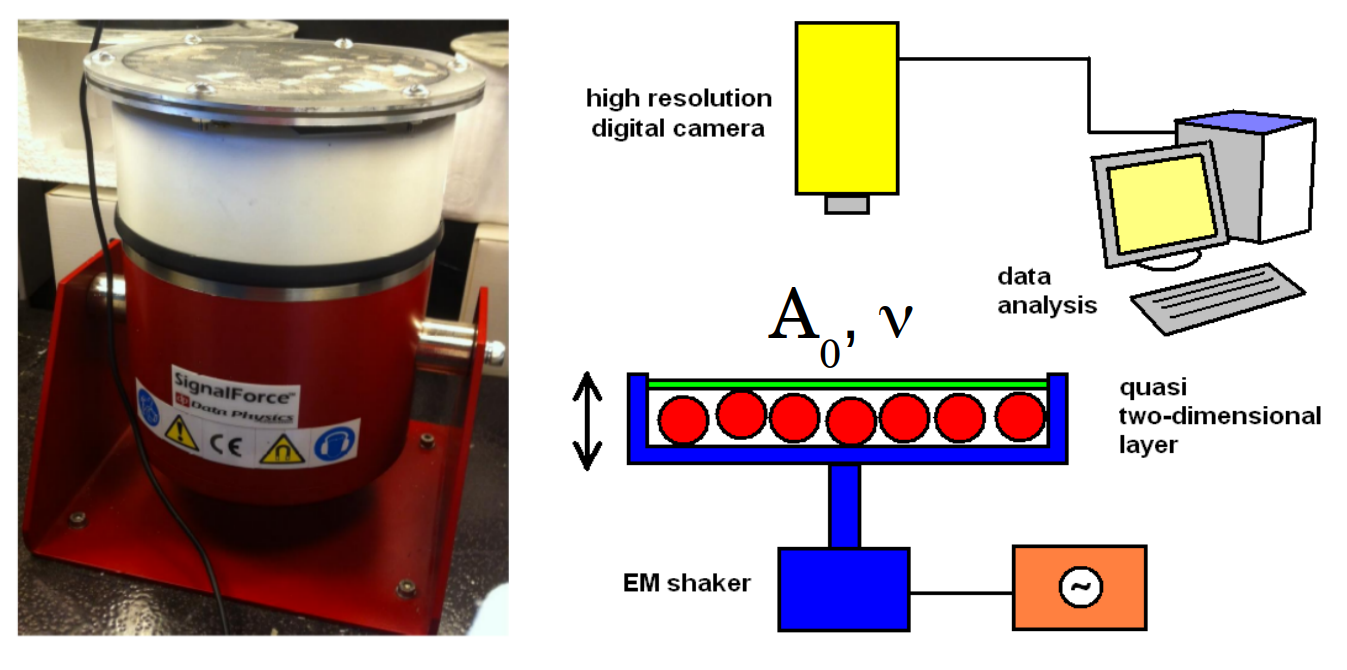}
		\caption{Schematic of the experimental setup.}
        \label{fig1m}
        \end{center}
\end{figure}
These similarities are very remarkable, given the essentially different physics that operates
in systems of granular particles as opposed to systems in thermal equilibrium. It seems that
collections of dissipative particles in highly-packed arrangements, subject to the continuous
injection of energy via aggitation, may reach steady-state configurations that can be understood in
terms of entropic arguments, with a minimal contribution from nonequilibrium effects, which can be
largely suppressed if experimental parameters are chosen adequately (these effects can be quantified
as a slightly larger tendency of particles to form long-lived local arrangements \cite{Miguel}). 
Therefore, as shown in Section \ref{theory}, entropy alone can explain the existence 
of orientational ordering in high-density fluid monolayers of granular rods under vibration.

Fig. \ref{fig1m} is a schematic of the experimental setup. Steel particles 1.2 mm in width and
an aspect ratio of 4 are confined
into a circular cavity 16 cm in diameter and 2 mm in height. The system is quasi-2D since
particles cannot pass over each other. The set is vibrated in the vertical direction with a frequency
$\nu$ close to 100 Hz, and an amplitude $A_0$ which is adjusted so that the average acceleration 
$\Gamma=A_0\nu^2/g$ is in the range 2-3, where $g$ is the acceleration of gravity. 
With these settings, and at the packing-fraction regimes investigated, the local
particle density is quite uniform and no large fluctuations can be detected. Spontaneous global rotation
of the whole sample is sometimes excited but the analysis of the results subtract this mode.
\begin{figure}[h]
        \begin{center}
	\includegraphics[width=3.in]{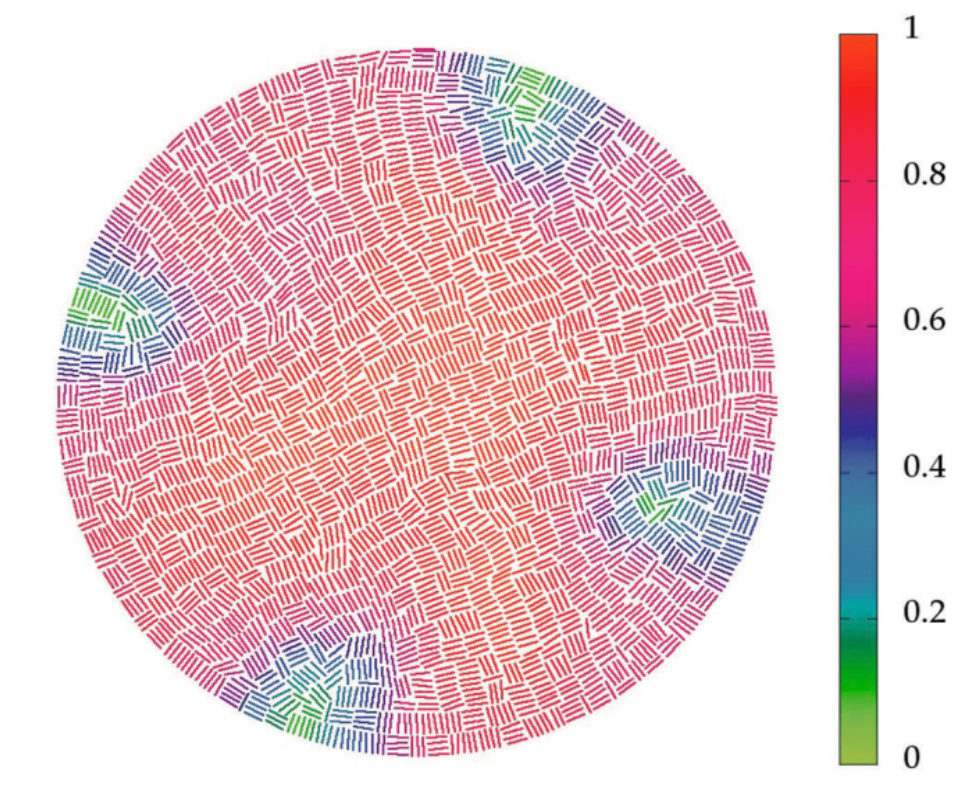}
		\caption{Colour map of the $Q_4$ order parameter of a granular monolayer of cylinders
		of aspect ratio 4 at packing fraction 
		$\eta=0.70$. A global 4-atic symmetry is apparent, but with four point defects which are clearly
		visible at regions with a depleted order parameter.}
        \label{fig_0}
        \end{center}
\end{figure}
As an example, consider Fig. \ref{fig_0}, 
which shows the local $4$-atic order parameter $Q_4$ in a monolayer
with packing fraction $\eta\simeq 0.7$. The local ordering field can be defined in terms of order
parameters, based on Eqn. (\ref{order_parameters}) but calculated locally over particles inside a 
small circular region. The
local nematic director is calculated by diagonalising the local order tensor, and all the order
parameters $Q_{2n}$ can be obtained from this reference direction. In this configuration a high value
of $Q_4$ can be seen throughout most of the system (on the contrary, the value of $Q_2$, not shown, is very low).

As mentioned above, the similarities between dissipative and equilibrium systems are not based 
simply on the observation that order parameters can be defined and that orientational ordering 
appears more or less abruptly as density is increased. An ordering local field can also be defined
in the system. When subject to conflicting boundaries with symmetries different from the inherent
symmetries of the configurations, point defects, where the orientational field can no longer
be defined, are excited. Fig. \ref{fig_0} shows a typical example where a globally $4$-atic configuration
of confined rectangles (projected steel cylinders) exhibits fours point defects located close to the cavity
wall, forming a square. This is the expected configuration of repelling defects, and leads to the
conclusion that there is an elastic field through which defects interact at long distances.
An elastic analogy may help elucidate the nature of these point defects. The Frank elastic free energy is
\begin{eqnarray}
	F_{\rm el} = \frac{1}{2}\int_A d{\bm r} \left[K_1\left(\nabla\cdot\hat{\bm n}\right)^2+
	K_3\left|\hat{\bm n}\times\left(\nabla\times\hat{\bm n}\right)\right|^2\right],
	\label{Frank}
\end{eqnarray}
where $\hat{\bm n}$ is the nematic field (director), and $K_1,K_3$ the elastic moduli associated to splay and
bend, respectively. In 2D nematics the twist mode is obviously absent. Note that in a $4$-atic phase
the two equivalent directors are orthogonal, $\hat{\bm n}\perp\hat{\bm m}$, so that the splay mode of
one corresponds to the bend mode of the other, and Eqn. (\ref{Frank}) is closed with respect to all
possible modes. This results, based on symmetry grounds, implies that $K_1=K_3\equiv K$, so that 
the elastic free energy can be written as 
\begin{eqnarray}
F_{\rm el} = \frac{1}{2}K\int_A d{\bm r} \left|\nabla\theta\right|^2,
\end{eqnarray}
where ${\bm n}=(\cos{\theta},\sin{\theta})$, and $\theta$ is the local orientation of the director.
This is the free energy of the $xy$ model, which is known to contain vortices as elementary
excitations. Thus, the equilibrium configuration of $\theta({\bm r})$ obeys Laplace's equation 
(harmonic solutions). The elementary excitation of this model, spin waves and local vortices or
topological defects, can be created spontaneously through thermal fluctuations, but defects 
can be driven by geometrical constraints. This is the case in
our granular monolayers. 

The nature of the defects can be obtained by examining the elastic field
in the neighbourhood of the defects. These are described in terms of a phase $\theta_0$ and a winding 
number $k$. Far from the defect the harmonic field is $\theta({\bm r}) = \theta_0+k\varphi$, 
where ${\bm r}=r(\cos{\varphi},\sin{\varphi})$. According to topology theory, there is a relation
between the p-rotational symmetry of the phase and the winding number, given by the topological charge $q$ of the defect:
\begin{eqnarray}
	q=\frac{2\pi w}{2\pi/p}=wp
\end{eqnarray}
In addition, Euler's theorem implies that there is a conservation of the
total topological charge, 
\begin{eqnarray}
	q_t=\displaystyle\sum_iq_i=p\chi,
	\label{Euler}
\end{eqnarray}
where the sum
extends over all defects in the system and $\chi$ is the Euler characteristics
of the container. For a disc (circular cavity) or a square, $\chi=1$; for
an annulus, $\chi=0$, and so on.\\
\begin{figure}
        \begin{center}
	\includegraphics[width=5.in]{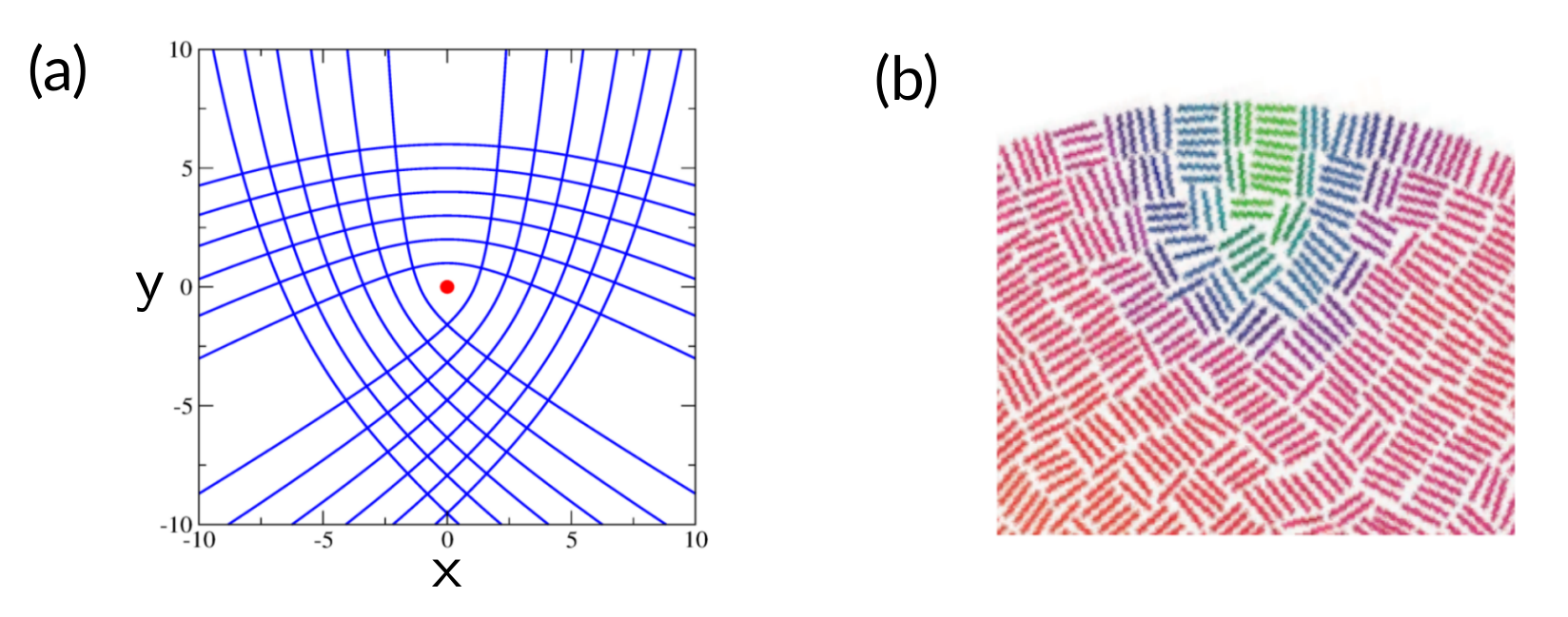}
		\caption{(a) Director field in the neighbourhood of a point defect with winding number $w=+1/4$.
		(b) Particle configurations in the region surrounding a point defect. Colour indicates
		the value of $Q_4$. A closed loop surrounding the defect has been drawn. The triangular symmetry
		of this loop (with curved sides) clearly matches the one resulting from the elastic field in panel
		(a), which indicates that the defect in the granular monolayer corresponds to a winding number $w=+1/4$.}
        \label{fig_1}
        \end{center}
\end{figure}
\\
Fig. \ref{fig_1} (a) shows a solution for the elastic field, in the neighbourhood of
a singularity of the field, using a winding number $w=+1/4$,
corresponding to a charge $q=1$ ($p=4$ for the tetratic). In panel (b) 
a real configuration of the particles in a defected region is shown. The
local orientation of the particles has been sketched. Clearly the 
overall symmetry of both fields, theoretical and experimental, coincide, which
leads one to conclude that the circular 
cavity is actually generating defects with
topological charge $+1$. In view of Eqn. (\ref{Euler}), it is immmediately
clear that topology predicts four $+1$ defects in the circular cavity, and
this is indeed the case. Note that other possibilities, with other types
of defects of different charge, could also exist in general; in this case
four $+1$ point defects are the only valid solution, as observed in the
experiment.
\begin{figure}
        \begin{center}
	\includegraphics[width=5.in]{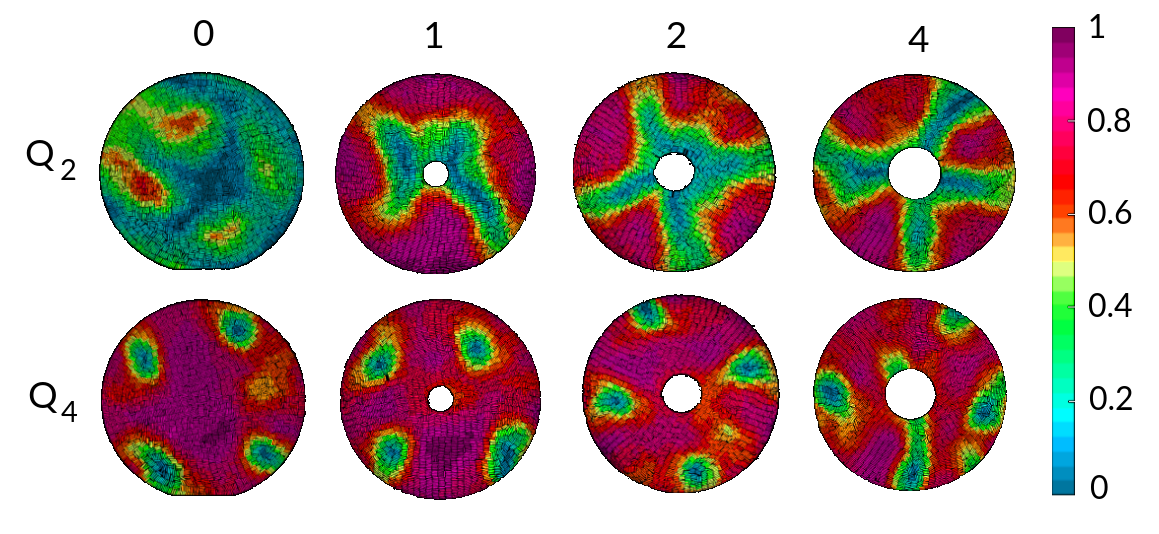}
		\caption{Colour maps of order parameters $Q_{2n}$, with $n=1$, $2$ and $4$.
		Columns are labelled by the numbers 0, 1, 2, 4, which correspond to the size (in cm) of the
		central obstacle in the cavity.}
        \label{fig_2}
        \end{center}
\end{figure}
We have looked at a more complicated cavity, the annulus. Here, two
concentric circular walls confine the particles, one with positive 
curvature and the other with negative curvature. Particles confined in this type
of containers have been examined before \cite{Garlea}. However,
we must keep in mind that topology predictions rely on the continuum
approximation. In our case this means that particle size should be much
less than any typical scale in the container. In the case of the circular
cavity, the ratio $R/l$, where $R$ is the cavity radius and $l$ the
length of the particle, is $\sim 20$. For the annulus we should focus on
the inner wall, $R_{\rm in}$. In Fig. \ref{fig_2} we show maps of the different
order parameters for different values of $R_{\rm in}$: 5, 10 and 20 mm.
This implies values $R_{\rm in}/l=1.25$, $2.5$ and $5$, which may be too
small for the continuum approximation to be valid . 
The predictions of topology theory may thus be compromised. We
will see that this is actually the case.

For an annulus $\chi=0$, and Eqn. (\ref{Euler}) predicts a total topological
charge of $0$. This does not mean necessarily that there are no defects
in the system, but that the {\it total} charge should be zero. Therefore,
it is obvious that a solution for the elastic field with only 
distortion, and no defects, is a possibility. This is presumably the solution 
with the lowest free energy. However, this is not what happens.
Instead, a complex pattern is excited. The pattern consists of smectic regions,
characterised by a high value of $Q_2$, separated
by regions of $4$-atic order containing a point defect. Up to four smectic regions can
be seen in Fig. \ref{fig_2}. The two types of regions are separated by domain walls.
The system without obstruction (left column) does not exhibit these structures.
In small nematic samples subject to distortion domain walls can be excited
(see e.g. \cite{Daniel-Velasco}). These structures are not contemplated by the simple
elastic theory supplemented by point defect, but might result by
extending elastic theory to incorporate domain walls, or $D-1$-dimensional
defects, where $D$ is the dimension of physical space. Anyway the 
continuum limit may be broken down in this case and more detailed analysis 
would be required to exlain the complex structures shown in Fig. \ref{fig_2}.

Many other confining geometries and particle shapes can be explored with the present setting.
For example, it would be interesting to explore particles with triangular shape, which may present
different types of inherent order. As shown by our theoretical DFT calculations, equilateral triangles 
stabilise a p-atic phase with $p=6$. A configuration with six $q=+1$ defects is expected in a cavity
with $\chi=1$. But there may be differences as regards not only how these defects appear in the cavity, but
also the number and values of the defect charges, as the only requirement involves the total charge
in the cavity. Another interesting system will be a fluid of right-angles triangles. As discussed in 
Section \ref{theory}, clustering effects may give rise to strong 8-atic correlations in the fluid
through the formation of relatively stable compact arrangements of particles which generate orientational
order along complementary directions and bulk phases with secondary directors. Whether this feature may 
or may not affect the number of nature of the defects remains to be seen.
Also, since clustering is expected to be stronger in the granular system than in a similar
system in thermal equilibrium, the cluster distribution and the fractions of different types
of clusters may change substantially in experimental granular monolayers subject to vibration.

\section{Conclusion}
\label{conclusion}

In this work we have summarised the current status of our own investigations on the ordering 
properties and liquid-crystalline phases of 2D hard particles of polygonal shapes.
Our approach is based on the interplay between two views: an equilibrium view, obtained by using
our standard DFT functional based on the SPT approximation and extensions, together with computer
simulations, and a nonequilibrium view,
based on experiments on vibrated monolayers of quasi-two dimensional particles. These two systems,
although very different from a physical point of view, seem to meet at some point in terms of
ordering behaviour and response to external fields. Experiments can
be used to verify the predictions of the theory, while the theory can suggest new experiments which
can be useful to test these predictions. 

In general, DFT predictions are qualitatively correct,
in that the symmetry of the phases and the overall region of stability with respect to density and
particle sizes are reasonably predicted, with the agreement even improving when three-body correlations
are included. However, we have examined a case where DFT fails dramatically: 
the case of right-angled triangles. In this system simulations indicate a strong tendency for the system to develop 
$4$-atic and $8$-atic correlations, due to the strong tendency of particles to locally arrange in dimers 
(of square, triangular and rhomboidal shapes) and tetramers (of square shape).
As a result, the symmetry of the bulk phase may be different from the one
dictated by the geometry of the particles. It then seems that clustering may be very important
for some nonregular polygonal shapes, giving rise to nontrivial symmetries than cannot be 
predicted by the standard DFT theories based on two-, or even three-body correlations. 

Based on the above findings, two future lines of research can be identified. On the one hand, improved DFT theories,
incorporating particle clustering in a consistent way (not as imposed static clusters with 
predefined fractions as in Section \ref{mixtures}), may be worthwhile to develop. In parallel, and view of the success of 
experiments based on vibrated monolayers, new granular particles can be explored in different
geometries, in an effort to extract information on the symmetries of the system by looking at
local order parameters and at the elementary localised excitations that result when that inherent
symmetries are subject to geometric frustration through confinement by different types of cavities.

\acknowledgments
Financial support from Grant No. PGC2018-096606-B-I00 (MCIU/AEI/FEDER,UE) is acknowledged.

\end{document}